\documentclass[fleqn,usenatbib]{mnras}

\usepackage{newtxtext,newtxmath}

\usepackage[T1]{fontenc}

\DeclareRobustCommand{\VAN}[3]{#2}
\let\VANthebibliography\thebibliography
\def\thebibliography{\DeclareRobustCommand{\VAN}[3]{##3}\VANthebibliography}

\usepackage{graphicx}	
\usepackage{amsmath}	
\usepackage{multirow}   
\usepackage{tabularx}
\usepackage{tikz}       
\usetikzlibrary{shapes.geometric, arrows.meta, positioning}
\usepackage{float}      

\tikzstyle{startstop} = [rectangle, rounded corners, minimum width=3cm, minimum height=1cm,text centered, text width = 3cm, draw=black, fill=red!30]
\tikzstyle{io} = [trapezium, trapezium left angle=70, trapezium right angle=110, minimum width=3cm, minimum height=1cm, text centered, text width = 3cm, draw=black, fill=blue!30]
\tikzstyle{process} = [rectangle, minimum width=3cm, minimum height=1cm, text centered, text width = 3cm, draw=black, fill=orange!30]
\tikzstyle{decision} = [diamond, minimum width=3cm, minimum height=1cm, text centered, text width = 3cm, draw=black, fill=green!30]
\tikzstyle{arrow} = [thick,->,>=stealth]

\title{RadioSED I: Bayesian inference of radio SEDs from inhomogeneous surveys}

\author[E.F. Kerrison et al.]{
Emily F. Kerrison,$^{1,2,3}$\thanks{E-mail: emily.kerrison@sydney.edu.au},
James R. Allison$,^{4}$
Vanessa A. Moss,$^{3,1}$
Elaine M. Sadler,$^{1,2,3}$
and Glen A. Rees$^{5}$
\\
$^{1}$Sydney Institute for Astronomy, School of Physics A28, University of Sydney, NSW 2006, Australia \\
$^{2}$ARC Centre of Excellence for All Sky Astrophysics in 3 Dimensions (ASTRO 3D) \\
$^{3}$ATNF, CSIRO Space and Astronomy, PO Box 76, Epping, NSW 1710, Australia \\
$^{4}$First Light Fusion Ltd., Unit 9/10 Oxford Pioneer Park, Mead Road, Yarnton, Kidlington OX5 1QU, UK\\
$^{5}$Freelance, Australia\\
}

\date{Accepted XXX. Received YYY; in original form ZZZ}

\pubyear{2024}

\begin{document}
\label{firstpage}
\pagerange{\pageref{firstpage}--\pageref{lastpage}}
\maketitle

\begin{abstract}
We present here \textsc{RadioSED}, a Bayesian inference framework tailored to modelling and classifying broadband radio spectral energy distributions (SEDs) using only data from publicly-released, large-area surveys. We outline the functionality of \textsc{RadioSED}, with its focus on broadband radio emissions which can trace kiloparsec-scale absorption within both the radio jets and the circumgalactic medium of Active Galactic Nuclei (AGN). In particular, we discuss the capability of \textsc{RadioSED} to advance our understanding of AGN physics and composition within youngest and most compact sources, for which high resolution imaging is often unavailable. These young radio AGN typically manifest as peaked spectrum (PS) sources which, before \textsc{RadioSED}, were difficult to identify owing to the large, broadband frequency coverage typically required, and yet they provide an invaluable environment for understanding AGN evolution and feedback. We discuss the implementation details of \textsc{RadioSED}, and we validate our approach against both synthetic and observational data. Since the surveys used are drawn from multiple epochs of observation, we also consider the output from \textsc{RadioSED} in the context of AGN variability. Finally, we show that \textsc{RadioSED} recovers the expected SED shapes for a selection of well-characterised radio sources from the literature, and we discuss avenues for further study of these and other sources using radio SED fitting as a starting point. The scalability and modularity of this framework make it an exciting tool for multiwavelength astronomers as next-generation telescopes begin several all-sky surveys. Accordingly, we make the code for \textsc{RadioSED}, which is written in \textsc{python}, available on Github.
\end{abstract}




\section{Introduction}

 Over the last few decades, advances in both observational and theoretical astronomy have brought great progress in our understanding of galaxy structure and evolution, as well as in our understanding of complex AGN-galaxy interactions. On the theoretical side, ever-increasing computational efficiency has allowed for more physically realistic simulations from the cosmological scale down to the hydro-dynamics of individual AGN jets (e.g. \citealt{Somerville2015PhysicalFramework}). On the observational side and amongst many other advances, large, uniform surveys spanning both optical (e.g. GAMA: \cite{Driver2009}, SDSS: \cite{YorkSDSS2000}, DESI: \cite{Dey2019}) and infrared wavelengths (e.g. WISE: \cite{Wright2010}) have provided millions of low-resolution, photometric measurements of galaxies across cosmic time. One place in which the two fields meet is spectral energy distribution (SED) fitting, where these broadband, photometric measurements are mapped onto theoretical models of galaxy structure and evolution to tell us about everything from star formation history, to the state and quantity of the dust and gas of a galaxy \citep{Conroy2013ModelingGalaxies}. However in the radio regime, where AGN emissions dominate the sky above a few milliJansky \citep{Windhorst1999, Richards1999}, SED fitting and analysis is still somewhat in its infancy.

 There are many SED fitting tools on offer for the interested astronomer, and \cite{Pacifici2023TheTechniques} provide an excellent summary of these to date. Yet while the library of publicly-available SED fitting codes has now grown to more than a dozen, fewer than half are capable of handling AGN emissions (as opposed to those from the host galaxy), and fewer still extend their SED fitting down into the radio regime. Amongst the most popular fitting codes, both CIGALE \citep{Burgarella2005, Boquien2019} and MAGPHYS \citep{daCunha2008} implement radio SED fitting, but only CIGALE implements a radio AGN component, which appears there in the form of a power law relationship between radio luminosity $L_{AGN}$ and radio frequency $\nu$, and serves mainly to correct for an apparent excess in the radio flux predicted due to star formation (see \citealt{Yang2022}, Section 5 for further discussion).
 
 Part of this lack of radio SED fitting tools is undoubtedly due to the relative sparsity of radio flux density measurements compared to those in the optical and infrared. Indeed until very recently, large-area radio surveys were limited to only a handful of frequencies; the NRAO VLA Sky Survey (NVSS; \citealt{Condon1998}) at 1.4\,GHz in the north, and the Sydney University Molonglo Sky Survey (SUMSS; \citealt{Mauch2006}) at 843\,MHz in the south are perhaps two of the most well-known.

 In recent years though this has begun to change, starting with the low frequency GaLactic and Extragalactic All-sky Murchison Widefield Array (GLEAM) survey \citep{Hurley-Walker2017} which surveyed the full southern equatorial sky with a 160\,MHz bandwidth in a previously under-explored frequency space below $300$\,MHz. This led to the first systematic studies of radio spectral shape at those frequencies by \cite{Callingham2017} and \cite{Ross2021}, where the former produced, amongst other things, the largest sample of ``Peaked Spectrum'' (PS) AGN to date, a particularly interesting class thought to represent the very youngest radio AGN \citep{ODea2021}, and the later studied their spectral variability. Indeed these works reinforced earlier findings (e.g. \citealt{ODea1998, Snellen2000, EdwardsTingay2004} to name but a few), that a small but scientifically significant fraction of radio AGN exhibit diverse radio spectral shapes that cannot be captured by a simple power law model alone.

Fortunately, we are on the cusp of many releases from extremely sensitive all-sky surveys at higher frequencies from new and upgraded instruments alike that will make such analyses possible at higher frequencies, including the Very Large Array (VLA, with the VLA Sky Survey VLASS; \citealt{Gordon2021ASurvey}), the Giant Metrewave Radio Telescope (GMRT, producing the TIFR GMRT Sky Survey TGSS; \citealt{Intema2017TheADR1}), the Australian SKA Pathfinder (both the Rapid ASKAP Continuum Survey RACS; \citealt{McConnell2020, Hale2021, Duchesne2023} and Evolutionary Map of the Universe EMU; \citealt{Norris2011}) and LOFAR (with the LOFAR Two-metre Sky Survey LoTSS; \citealt{Shimwell2017}). Yet despite the large bandwidth of many of these instruments, no single survey alone will be enough to characterise broadband radio spectra spanning 700\,MHz through to tens or even hundreds of gigahertz. This is especially so since the standard practice for radio catalogues is to compress the observed bandwidth into a single flux density measurement at some central frequency. Therefore any approach to radio SED fitting on population-scales will still require a careful combination of several, independent radio surveys.

It is timely, therefore, that we present here \textsc{RadioSED}\footnote{The code is made available to the user on Github:\url{https://github.com/ekerrison/RadioSED/}}, a framework for extending the low-frequency, all-sky AGN spectral modelling of \cite{Callingham2017} by compiling SEDs from all suitable, large-area radio surveys to date. This framework grew out of a desire to identify and characterise a large, uniform sample of PS AGN with spectral peaks spanning a few megahertz through to tens or even hundreds of gigahertz, a task which was identified as a key area for future research in the review of \cite{ODea2021}. However by construction, the model classification performed by \textsc{RadioSED} allows for the characterisation of both the young, PS AGN which were the main focus of \cite{Callingham2017}, as well other radio AGN with flat, steep, and inverted SEDs, representing beamed blazar-like emission, larger (likely evolved PS) radio sources, and extremely young PS sources respectively, so that this framework has the potential to better inform us about the broadband spectral properties of the radio AGN population as a whole. For this reason our SED construction is complementary to that provided by existing tools like SPECFIND which perform more complex, algorithmic cross-matching between surveys but focus on sources with a power law spectral shape \citep{Vollmer2005TheCatalogues, Stein2021TheFrequencies}. Indeed at present, the models implemented in \textsc{RadioSED} are designed to differentiate between spectral shapes whilst being broadly applicable to different radio AGN environments, with a particular focus on identifying the young, PS sources. However, the modularity of our framework means that additional models incorporating more physical subtleties, such as differentiating between synchrotron self-absorption and free-free absorption of the radio jets, could be easily incorporated in the future. Not only that, but the automated and modular nature of this framework means it can be easily extended to include new surveys as they are released to the public, covering a wider sky area, a different frequency range or both at a greater sensitivity than ever before. The output from \textsc{RadioSED}, when considered alongside the output from semi-analytical models of radio jets like those of \cite{Turner2015ENERGETICSNUCLEI}, further developed in \cite{Turner2023RAiSE:Lobes}, will also allow for a deeper analysis of broadband radio spectral properties akin to what is already possible with optical and infrared SED analysis.

In Section~\ref{sec:sed_construction} we outline the process of compiling individual SEDs from large area surveys observed at different epochs. Section~\ref{sec:bayesian_modelling} presents our Bayesian approach to modelling and classifying the SEDs of individual radio sources, and in Section~\ref{sec:radiosed_verification} we discuss the accuracy and reliability of this modelling with reference to both synthetic data, and observational data for a selection of sources with well-known SED shapes. A summary of this work, and plans for future science applications of \textsc{RadioSED} are discussed in Section~\ref{sec:conclusions}, while a second, forthcoming paper in this series will present a deeper scientific analysis of the sources in our chosen pilot field, Stripe 82 \citep{Abazajian2009}.

\section{SED construction}\label{sec:sed_construction}

\begin{figure}
	\includegraphics[width=\columnwidth]{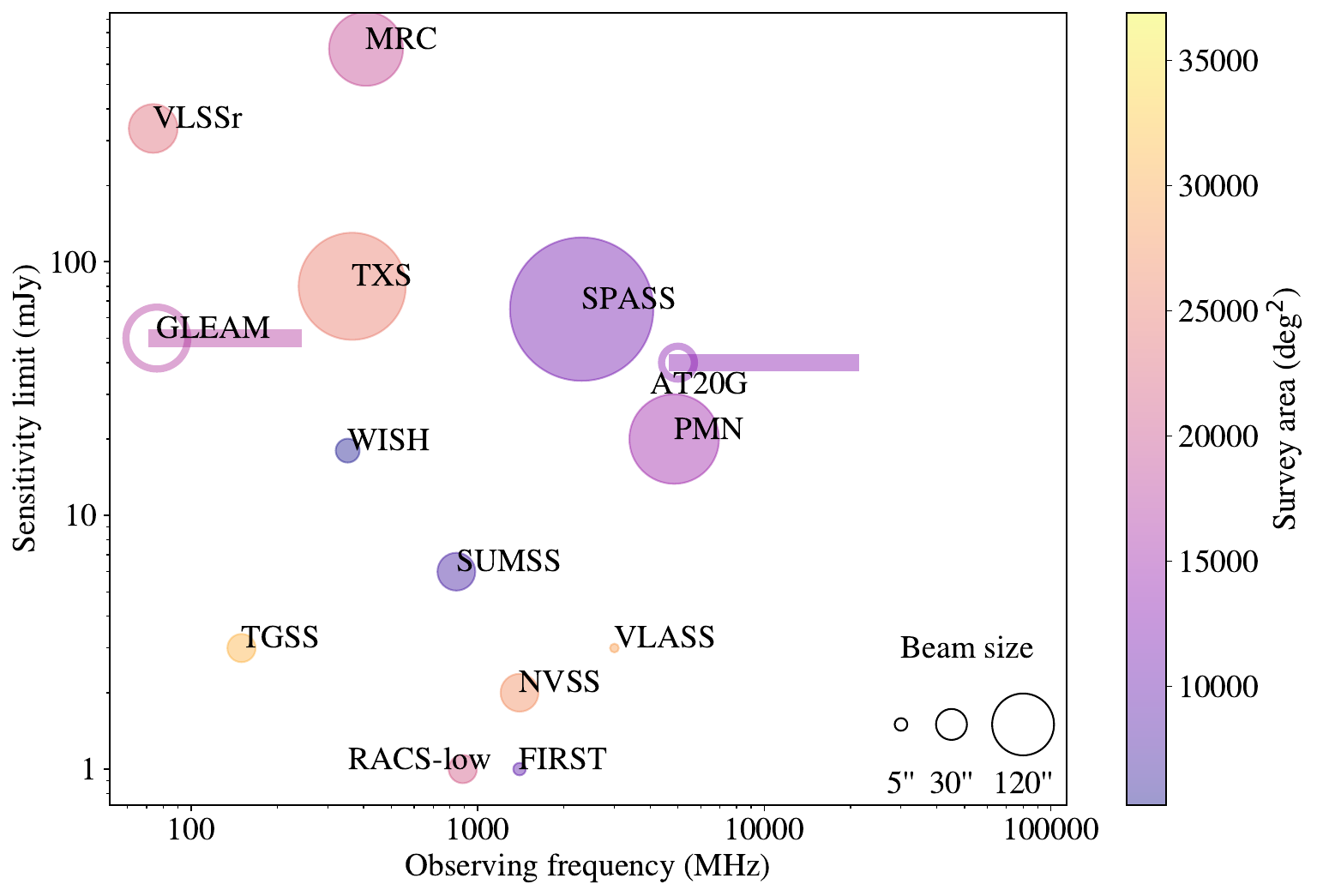}
    \caption{Frequency coverage of radio surveys to date which have observed more than $\sim$10$^2$ square degrees of southern equatorial sky. Rectangular regions indicate the maximum frequency range of broadband surveys, which report flux density measurements at multiple frequencies in their respective catalogues. Surveys with measurements at a single frequency are indicated as circles, with the size of the circle indicative of beam size and the colour indicative of survey area.}
    \label{fig:southern_radio_surveys}
\end{figure}

In order to identify interesting classes of radio AGN by their spectral shape, we must first construct their broadband SEDs in a reliable and reproducible manner. Ideally, these SEDs should cover a sufficiently large range of frequencies so as to identify spectral peaks which may manifest anywhere from a few hundred megahertz through to the gigahertz regime. Doing so will allow us in particular to capture a representative sample of the peaked spectrum population spanning a range of both ages and redshifts (recalling that the frequency of the spectral peak is inversely proportional to the linear size - and hence the age - of a source, but that this peak may be lowered in the observer's frame due to cosmological redshift; \citealt{Snellen2000, deVries2009, Jeyakumar2016}). To date, the southern equatorial sky has been relatively well sampled in frequency space, as outlined in Figure~\ref{fig:southern_radio_surveys}, though the sensitivity and epoch of this coverage varies wildly. Since no single survey or instrument is yet capable of achieving comparatively broad coverage, we must combine data from multiple surveys, instruments and even epochs to achieve our stated goals.

\begin{table*}
\begin{tabular}{m{14mm}m{13.7mm}m{7mm}m{7.3mm}m{7mm}m{10.7mm}m{8mm}m{12mm}m{11.5mm}m{7mm}m{30mm}} 
\hline
Name & Frequency & Beam & n$_{\text{sources}}$ & Area & Instrument & Flux & Flux limit & Final Obs. & r$_\text{match}$ & Reference\\
& & FWHM & & & & limit & Frequency & Date & 95\% & \\
& (GHz) & (arcsec) & & (deg$^2$) & & (mJy) & (GHz) & & ('') &  \\
\hline \hline
VLSSr  & 0.0738   & 75   & 92964    & 30939    & VLA & 335   & 0.074 & 2007 & 25  & \cite{Lane2014TheVLSSr}   \\
TGSS   & 0.15    & 25   & 623604   & 36900    & GMRT    & 3  & 0.15    & 2012 & 18  & \cite{Intema2017TheADR1}  \\
CCA    & 0.080-0.16 & 222  & 2213 & N/A   & CCA   & 1200  & 0.16    & 1984  & 43  & \cite{Slee1995RadioArray.} \\
GLEAM  & 0.076-0.227 & 120  & 307455   & 24831    & MWA & 50    & 0.2    & 2014  & 25  & \cite{Hurley-Walker2017}   \\
TXS    & 0.365    & 360  & 66841    & 31538    & Texas Int.  & 80    & 0.365    & 1983  & 16  & \cite{Douglas1996TheMHz}     \\
MRC    & 0.408    & 172  & 12141    & 25770    & Molongolo & 690   & 0.408    & 1974 & 22  & \cite{Large1981TheSources.}   \\
SUMSS  & 0.843    & 45   & 211050   & 8100 & MOST    & 6  & 0.843    & 2003 & 14  & \cite{Mauch2013}  \\
\textbf{RACS-low}   & \textbf{0.8875}  & \textbf{25}   & \textbf{2462639}  & \textbf{28020}    & \textbf{ASKAP}   & \textbf{1}  & \textbf{0.8875}  & \textbf{2020}   & \textbf{N/A}  & \textbf{\cite{Hale2021}}   \\
NVSS   & 1.4   & 45   & 1773484  & 33827    & VLA & 2  & 1.4   & 1997 & 11  & \cite{Condon1998} \\
SPASS  & 2.307   & 645  & 23389    & 16000    & Parkes  & 65    & 2.307   & 2010 & 48  & \cite{Meyers2017AData}     \\
PMN    & 4.850   & 252  & 50841    & 23265    & Parkes  & 20    & 4.850   & 1990  & 43  & \cite{Griffith1993TheReduction}     \\
Randall11  & 1.4-8.6  & 40   & 42   & N/A   & ATCA    & 1500  & 2.7   & 2008 & 129    &  \cite{Randall2012}   \\
ATPMN  & 4.8-8.6  & 2.8  & 9040 & 7757 & ATCA    & 7    & 4.85   & 1994  & 18  & \cite{McConnell2012ATPMN:Survey}   \\
Sajina11   & 4.86-90 & 16   & 159  & N/A   & VLA & 40    & 20  & 2010  & 19    & \cite{Sajina2011High-frequencyField} \\
AT20G  & 5-20 & 34   & 5867 & 20086    & ATCA    & 40    & 20  & 2008  & 19  & \cite{Murphy2010TheCatalogue}\\
Ricci06   & 18.5-22   & 120  & 250  & N/A   & ATCA    & 1000  & 5   & 2002   & 103    & \cite{Ricci2006High-frequencyQuasars} \\
ACC   & 85.8 - 373 & 115  & 3364  & N/A & ALMA & 1 & 95 & 2018 & 22  & \cite{Bonato2019ALMASources}    \\
\hline
VLSS   & 0.074  & 80   &  92965  & 30939 & VLA & 0.2 &  0.074 & 2007 & -- & \cite{Cohen2007TheSurvey}  \\
PKSCAT90  & 0.080-22   & 60   & 8264 & 5007 & Parkes  & 50    & 1.4   & 1990  & -- & \cite{Wright1990ParkesFacility.}  \\
WISH   & 0.352    & 18   & 90357    & 5252 & WRST    & 18    & 0.352    & 1998  &  -- & \cite{DeBreuck2002ASurvey} \\
FIRST  & 1.4   & 5    & 946432   & 10575    & VLA & 1  & 1.4   & 2011 & --  & \cite{Helfand2015TheIdentifications}   \\
VLASS  & 3   & 2.5  & 3381277  & 33880    & VLA & 3  & 3   & 2019 & -- & \cite{Gordon2021ASurvey}   \\
C-BASS & 5   & 2700 &  -- & 41,253 & C-BASS &  --  & -- & -- & -- & \cite{King2010TheData}\\
CRATES & 8.4 & -- & 11131 & N/A & various & 65    & 4.8 & 2006 & -- & \cite{Healey2007CRATES:Sources}    \\
Kühr & 0.012-89  & -- & 518 & N/A & various & -- & -- & 1980 & -- & \cite{Kuehr1981A5-GHZ} \\
\hline
\end{tabular}
\caption{A summary of all of the wide-field surveys considered as part of our Bayesian framework. Those in the top half of the table have been included in the final flux density tables for sources of interest, those below the line were excluded from fitting for reasons further elaborated in-text. RACS-low \citep{Hale2021} was used as the base catalogue to which all other catalogues were matched, and is emphasised in the table. Empty entries either indicate that the value could not be found in the literature, or varied substantially across the survey. Surveys with an area marked as "N/A" were targeted followup campaigns.}\label{tab:survey_info}
\end{table*}

Our approach to SED construction is outlined in the following sub-sections. In short, we rely on public datasets available via the VizieR catalogue access tool \citep{Ochsenbein2000}, combined with position-based cross-matching, to build up reliable SEDs containing flux density measurements of physically-associated sources observed at a range of frequencies. The minimal assumptions in our cross-matching algorithm are what sets this approach apart from the SPECFIND tool \citep{Vollmer2005TheCatalogues, Stein2021TheFrequencies}. Since SPECFIND assumes a spectral shape at the stage of cross-matching catalogues, the tool largely disregards flux density measurements which deviate from this expected trend as `waste', though some work on identifying PS sources peaking at specific frequencies has been done recently (see \cite{Stein2021TheFrequencies} Section 5). However our purely position-based matching algorithm keeps all flux density measurements within a certain distance threshold, allowing us to better capture the more complex spectral curvature of interesting AGN sub-classes, such as the peaked spectrum sources. The caveat to using such minimally-informed crossmatching is that there is a higher chance of either matching non-physically-associated flux densities at different frequencies, or of making only a partial match between complex sources resolved in some surveys and not others. These downsides are mitigated by using the survey with the equal-highest resolution (aside from  ATPMN and Sajina11 which are targeted surveys of point-like sources) as our reference survey, and by selecting a statistically-robust match radius determined individually for each survey pair, as outlined in Sections \ref{sec:included_surveys} and \ref{sec:crossmatching} respectively.

\subsection{Catalogues}\label{sec:included_surveys}

The first step in a uniform approach to broadband radio spectral modelling is to determine how to compile relevant flux density measurements and then, once compiled, to assign them to individual sources.

On the one hand, at the level of individual sources, we require flux density measurements across several decades in frequency space in order to construct reliable SEDs with which to identify spectral shape (including any peaks). On the other hand, at the sample level these flux density measurements should ideally be drawn from all-sky, or at least large area ($\geq 10^3$ deg$^2$) untargeted surveys in order to maximise both the reliability and completeness of the full sample. These two concerns, the reliability of individual SEDs, and the completeness of the full sample, must be balanced against each other in determining how to draw flux density measurements from the literature.

After a manual search of both the tables collated in \cite{Stein2021TheFrequencies} and those elsewhere in the literature, we created a list of 25 candidate surveys outlined in Table~\ref{tab:survey_info}. Of these, 17 were selected for final inclusion in our SED fitting, which can be found above the dividing line in said table.

Broadly speaking, we favoured radio surveys which fit the following criteria:

\begin{enumerate}
    \item Large area, untargeted surveys ($\geq$ 5,000 deg$^2$) to minimise selection bias.
    \item Recent observation (since 2000) to minimise the effects of blazar variability in multi-epoch SEDs and flux scale inconsistencies, especially since the Australia Telescope Compact Array (ATCA) flux scale was revised in 1994 to more closely align it with that of the VLA, and many Southern equatorial surveys have been conducted with the ATCA \citep{Reynolds1994}.
    \item Observing frequency between 30\,MHz-100\,GHz to capture a broad range of spectral peaks, and hence a peaked spectrum sample covering a range of ages and redshifts.
\end{enumerate}

Where large frequency bands lack surveys fulfilling all 3 criteria, we have occasionally chosen to include data from surveys fulfilling only one of either the area or epoch criteria, in order to maximise spectral coverage across our sample (e.g. the inclusion of the MRC and TXS surveys around 400\,MHz). However, the modular nature of our fitting framework means that these surveys can be easily exchanged for more suitable ones as new data is made public.

From these surveys, a reference dataset must be selected upon which to base our SED construction. A closer consideration of all three criteria above led to the selection of RACS-low (\citealt{Hale2021}; emphasised in the table) for this purpose. Not only is it recent (observations date from 2020) and of an excellent sensitivity for our work (1\,mJy at $\sim 800$\,MHz), but it also covers the entire southern, equatorial sky at a positional accuracy of 1-2 arcseconds \citep{McConnell2020, Hale2021}. Therefore, for this work all other included surveys were matched against the RACS-low catalogue in order to build the SEDs for individual objects, and the positions of sources are given by their coordinates in the RACS-low catalogue.

In some cases, targeted follow-up campaigns of large area surveys at different frequencies have been included (e.g. \citealt{Ricci2006High-frequencyQuasars}), because they provide additional sky coverage at certain frequencies not well covered by the untargeted surveys already selected. These are indicated by a ``N/A'' value in the survey area column of Table~\ref{tab:survey_info}. We have also extended the frequency space explored up to the milimetre regime with the inclusion of the ALMA Calibrator Catalogue (ACC; \citealt{Bonato2019ALMASources}), to allow us to capture the very youngest peaked spectrum sources which turn over at tens of gigahertz or higher.

The surveys which were ultimately discarded were ruled out on the basis of flux scale errors (VLASS, VLSS; as documented in \cite{Gordon2021ASurvey} and \cite{Cohen2007TheSurvey} respectively), a highly irregular beam (WISH) or poor $uv$-coverage leading to missing flux on scales comparable to that of our reference catalogue (FIRST), or incomplete documentation of the observing campaigns from which flux densities were derived (PKSCAT90). We additionally ruled out the use of the CRATES data owing to a lack of information about resolution across the survey area, and discrepancies with the AT20G flux scale (see \citealt{Healey2007CRATES:Sources} section 3.3), and we likewise rule out the Kühr sample owing to the age of the catalogue and the assortment of observing parameters. 
Finally, we discard the C-BASS survey as its southern equatorial portion is not yet released, and its focus is on diffuse galactic emission.

\subsection{Catalogue Cross-matching}\label{sec:crossmatching}

Simple, position-based crossmatching against the reference catalogue (RACS-low) was performed for each catalogue in the top region of Table~\ref{tab:survey_info} using the CDS XMatch service \citep{Pineau2020ThePlans} accessed via Astroquery \citep{Ginsburg2019Astroquery:Python}. Matches were accepted as physically associated, and not a chance coincidence, if the separation between catalogued coordinates fell within the radius given in the $r_{\text{match}}$ column in arcseconds. This value varies across surveys since each catalogue used has a different source density dependent on its frequency, sensitivity and beam size. In each case, the radius was chosen using a simple statistical method based on the approach of \cite{Best2005ASurvey,Sadler2007, Ching2017} and others, who match radio sources with both real and randomly positioned optical counterparts to ensure the completeness and reliability of their real samples.

\begin{figure}
	\includegraphics[width=\columnwidth]{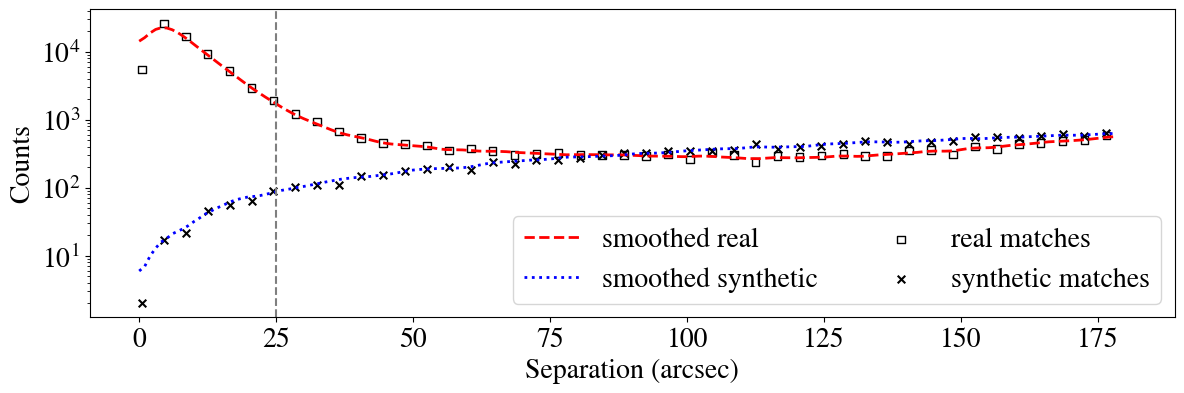}
    \caption{A representative histogram showing the separation between GLEAM sources \citep{Hurley-Walker2017} and their counterparts in the real RACS-low reference catalogue (unfilled squares), and also with their counterparts in the synthetic reference catalogue (filled crosses). The red and blue dashed curves are the counts smoothed. The vertical line indicates the 25 arcsecond separation threshold at which the synthetic cross-matches reach 5\% of the real cross-match count.}
    \label{fig:xmatch_hist}
\end{figure}

First, a `synthetic' catalogue was created as a counterpart to the reference catalogue by shifting the coordinates of every source by 1 degree at a random position angle. The magnitude of the shift was chosen to be large enough to remove the physical clustering produced by extended, multi-component sources, which \cite{Blake2003} demonstrated is significant out to only 0.3 degrees in both NVSS and FIRST. This resulting synthetic catalogue preserves the source density of the original, and can therefore be meaningfully compared with it. We then performed cross-matching against both the synthetic and real RACS-low catalogues for every other survey of interest, and examined the number of matches as a function of radius, as shown in Figure~\ref{fig:xmatch_hist}. We determine the maximum acceptable match radius to be the point at which matches with the synthetic catalogue comprise no more than 5\% of the number of matches with the real reference catalogue when the counts of both are smoothed across bins using a Gaussian filter. All crossmatches below this threshold are statistically very likely to represent flux density measurements of real, physical sources observed across the various surveys, and are hence included as reliable measurements in the SEDs of individual sources.

\subsection{Source Compactness and Complexity}\label{sec:compactness}

\begin{figure*}
    \includegraphics[height = 6.5cm]{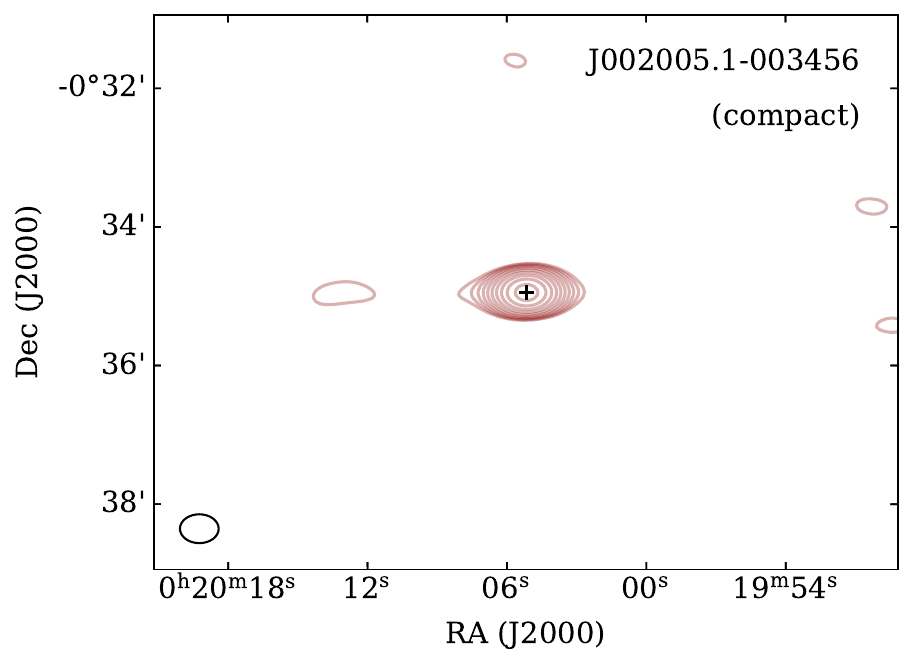}
    \includegraphics[height=6.5cm]{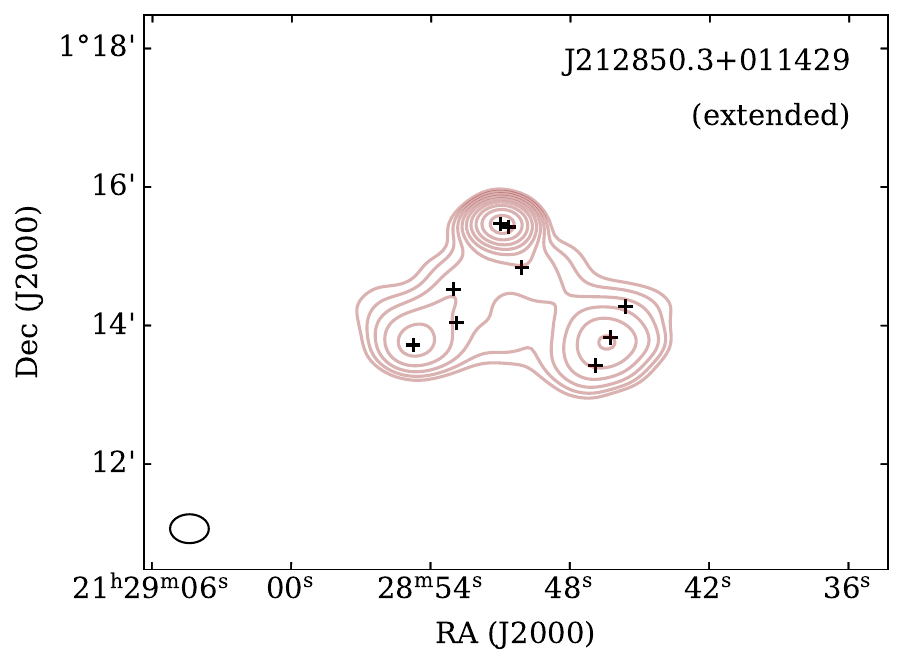}
    \caption{Examples of sources which appear compact (left) and extended (right) in the RACS-low observations. Red contours are logarithmically spaced between $5-200$\,mJy. Black crosses indicate the positions of the Gaussian components belonging to the island source displayed by the radio contours. The RACS-low beam size is given in the bottom left of each plot.}
    \label{fig:racs-contours}
\end{figure*}

Since all of our chosen surveys were observed at different resolutions, and many of them with interferometers, the morphology of a source has the potential to affect its SED shape. In general, radio sources can be broadly grouped into two morphological categories: compact, and extended. This classification depends on the resolution of the observations (since at some level, all radio sources must be extended), but it is nevertheless a useful distinction in the context of SED fitting, particularly for the identification of peaked spectrum sources, which are known to be extremely compact, with a linear diameter of ${\lesssim 1}$\,kpc at all observable radio frequencies \citep{ODea2021}. Sources of each kind as seen in RACS-low are displayed in Figure \ref{fig:racs-contours}. In terms of the relevance of source morphology for \textsc{RadioSED}, sources with extended structure may suffer from under-reported flux densities due to emission on scales larger than $ \lambda/2B_{min}$ being resolved out (by the Nyquist-Shannon sampling theorem), where $B_{min}$ is the shortest baseline length. While we do not attempt to correct for this in our SED fitting, we do check for the compactness of each source across 3 orders of magnitude in frequency space, so that extended sources which may suffer from missing flux can be flagged for later interpretation if desired. We note also that across the RACS-low catalogue, $\sim27\%$ of sources brighter than 10\,mJy are extended on scales which lead to multi-component Gaussian fits. Therefore, while the majority of sources fit by \textsc{RadioSED} will likely be compact, some significant fraction will be extended at gigahertz frequencies, requiring a more careful interpretation of their multifrequency SED. 

As stated, we obtained a measure of source compactness at three representative frequencies from across our selected surveys by utilising three surveys with the widest coverage: GLEAM \mbox{($\sim10^{-1}$\,GHz)}, RACS-low ($\sim10^{0}$\,GHz) and AT20G ($\sim10^{1}$\,GHz; \citealt{Murphy2010TheCatalogue}).

At the lowest frequencies, compactness on arcminute scales was obtained by taking the ratio of peak to integrated flux across the wide GLEAM band (170-231\,MHz), using the \texttt{Fpwide} and \texttt{Fintwide} catalogued values \citep{Hurley-Walker2017}. 

This process was repeated at just below 1\,GHz for RACS-low using the \texttt{Fpk} and \texttt{Ftot} catalogue entries which give an indication of compactness on arcsecond scales \citep{Hale2021}. For this survey, the number of Gaussian components into which each source was divided (reported as $N_g$ in the catalogue) was also recorded as an extra measure of source extent, since sources with complex, resolved morphology will be fit by multiple Gaussian components within the one island.

At high frequencies, flux-based measures of compactness in the image plane could not be used, since the AT20G survey only recorded peak fluxes and made use of only the short baselines in the ATCA's hybrid arrays \citep{Murphy2010TheCatalogue}. Instead, we made use of the additional AT20G High Angular Resolution Catalogue \citep{Chhetri2013TheCatalogue} which measured compactness on sub-arcsecond scales, reported as a visibility ratio, by utilising the raw 6-km visibilities from the original AT20G observing runs.

In all three cases, compactness is represented numerically on a linear scale from 0 - 1, with 1 being most compact. In the case of RACS-Low, we obtained our additional measure of extension from the component breakdown, where this time a value of 1 indicates the most compact sources (fit by 1 Gaussian component in the image plane), and anything higher indicates some extension (fit by multiple components). Since these measures of compactness cover three orders of magnitude in frequency space, they are a useful complement to the classifications returned from SED fitting alone. For example, if a source with an SED that turns over is compact in all three surveys, this is a good indication we have identified a true compact, peaked spectrum source. On the other hand, if a source has a compactness measure that is inversely proportional to frequency, that may be an indication of a compact core embedded in larger, extended lobes, which tend to manifest most clearly at low frequencies due to their steep spectral index \citep{Miley2008DistantEnvironments, Blandford2019RelativisticNuclei}. Thus these compactness measures provide a useful, if simplistic, indication of source morphology as it pertains especially to identifying peaked spectrum sources, and they do so without the need for high resolution imaging.

In future work, additional measures of source compactness may be easily incorporated into this framework. In particular, recent and ongoing observations of interplanetary scintillation made with both the MWA \citep{Morgan2022} and ASKAP \citep{Chhetri2023} will provide an invaluable measure of source compactness down to sub-arcsecond scales at frequencies at and below $\sim 1$\,GHz.

\subsection{Source Variability}\label{sec:variability}

\begin{figure*}
    \centering
    \includegraphics[width=\textwidth, trim={1cm 0cm 6.5cm 0},clip]{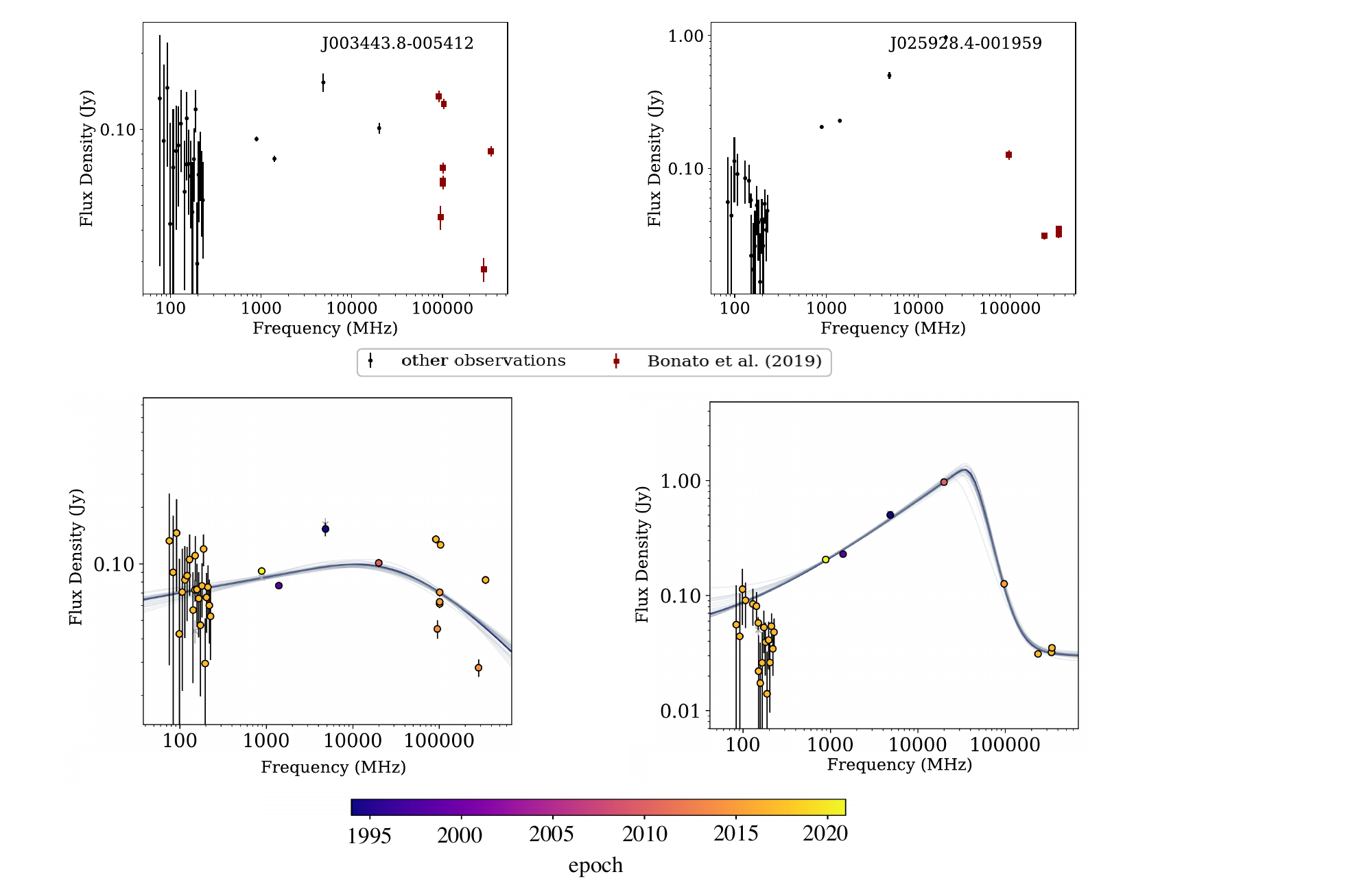}
    \caption{An example of a source exhibiting significant ($>10\sigma$) variability in the ALMA band (left), and one with a relatively stable flux density (right). In the top panel flux density measurements from the ALMA Calibrator Catalogue \citep{Bonato2019ALMASources} are indicated by dark red squares, black points indicate the other flux density measurements from the surveys in Table~\ref{tab:survey_info}. In the bottom panel, observations are coloured by observing epoch as drawn from the publication date of the respective survey in Table \ref{tab:survey_info}, with the model from \textsc{RadioSED} shown in grey. From a manual inspection of this representation, J003443.8-005412 (left) appears to exhibit more variability between sampled epochs than does J025928.4-001959 (right).}
    \label{fig:alma_variable}
\end{figure*}

The variability of radio AGN is a complex topic and no doubt our understanding is not yet complete, so we limit our discussion here to a consideration of the known variability mechanisms that might impact our SED fitting. 

AGN variability may be intrinsic, caused by the clumpiness of the circumnuclear material \citep[e.g.][on the clumpy torus]{Risalti2002}, a changing rate of accretion onto the supermassive black hole \citep{Czerny2009ACCRETIONSOURCES}, or by variations and instabilities within the radio jets themselves which may produce propagating shocks \citep[e.g.][]{Marscher1985, Kovalev2006Strong1997-2003, Hovatta2008Long-termCharacteristics}, and all of these may be beamed, increasing the degree of observed variability and producing a blazar \citep{Blandford2019RelativisticNuclei}.

Variability may also be extrinsic, caused by changing diffraction and refraction of light from the source as it passes through the turbulent interstellar medium of the Galaxy (Interstellar Scintillation - "ISS"; \citealt{Rickett1968, Rickett1984}) or the solar wind (Interplanetary Scintillation - "IPS"; \citealt{Clarke1964IPS} see also \citealt{Hewish1964InterplanetarySources}). All of these mechanisms may operate at once, producing complex variability across the radio spectrum on timescales from milliseconds to years, and in magnitude from a few percent, to several times the minimum recorded flux density \cite[e.g.][]{Torniainen2005LongCandidates, Ross2021}. For this work, our main interest is in how this variability could affect the SED shape of our sources when combining multi-epoch data, as well as in how transient spectral peaks could contaminate any eventual peaked spectrum source sample. We therefore care about the cadence at which variability occurs, the rate of variability within the radio AGN population, and the degree to which it occurs, as our sample is most likely to suffer from contamination or incompleteness when all three are extreme.

First of all, we must consider the temporal cadence at which we expect radio AGN to vary. Since IPS typically occurs on timescales of a few seconds in arcsecond-scale sources \citep{Hewish1964InterplanetarySources}, its effects will be averaged out by the integration time of the surveys used here; the ATCA has a standard integration cycle of 10 seconds\footnote{\href{https://www.narrabri.atnf.csiro.au/observing/users_guide/html/atug.html}{ATCA User Guide}}, and of the VLA surveys, VLSSr uses the shortest integration time at 10 seconds \citep{Cohen2007TheSurvey}. Other surveys here use comparable or longer integrations. By contrast, ISS fluctuations affecting the most compact sources happen on timescales of hours to days and also exhibit an annual cycle \citep{Rickett2001RadioScintillation, Bignall2003Rapid1257326}. Likewise, intrinsic variability has been observed over periods ranging from months to years at gigahertz frequencies \citep{Hovatta2007, Koay2018TheAGNs}. Therefore, while IPS will not have a meaningful impact on the flux densities used for our SEDs, both ISS and intrinsic variability may impact our propsosed SED analysis.

Fortunately, the rate and degree of radio variability within the compact, radio AGN population both appear to be relatively low, with quasar-type AGN exhibiting more variability than their galaxy-type counterparts \citep{Tinti2005HighSources}. \cite{Sadler2007} studied the variability of compact sources at 20\,GHz in an untargeted survey, finding a median variability of 6.9\% with only 5\% of sources exhibiting variability at the 30\% level or higher. Likewise at lower frequencies, \cite{Ross2021} examined variability in the GLEAM band below $\sim 300$\,MHz and found it to be low, with only $\sim$ 1.5\% of their full sample classed as variable, and of this variable subset only 16\% (51 out of 21,558 total sources) were found to exhibit extreme variability leading to changes in their spectral shape over the 3 year monitoring period. They did however note that peaked spectrum sources were over-represented in the variable subset, comprising about 30-40\% of this population, as were sources with high-frequency components. This is at odds with the definition of a peaked spectrum source as having a quiescent spectrum with variability below the 10\% level \citep{ODea1998}. However, in other followup campaigns of broadband instantaneous observations, a large degree of variability from apparent peaked spectrum sources has been shown to be the result of contamination of peaked spectrum samples by flaring blazars with a self-absorbed component, which produces a transient spectral peak in an otherwise flat radio spectrum \citep{Orienti2007, Torniainen2007, Mingaliev2012MultifrequencyTelescope}. 

To better address the question of blazar contamination in our fitting, as well as to provide additional frequency coverage, we have chosen to include the ALMA Calibrator Catalogue \citep{Bonato2019ALMASources} in our SEDs. This contains all calibrator observations made between 2011 and 2018 and at 90-370\,GHz, and many (though not all) of these sources are known blazars. In sources exhibiting variability in the ALMACAL catalogue (ACC) at the $10\sigma$ level, we do not use the flux density measurements when performing SED fitting, but we do flag the source as variable. We also return the maximum ALMA Variability Index (VI) as defined in \cite{AkritasBershady1996} and \cite{Barvainis2005} for all sources with more than 2 measurements in any given ALMA band. Individual sources exhibiting ALMA variability and a quiescent spectrum are shown in Figure~\ref{fig:alma_variable}, though we stress that this perceived variability (or lack thereof) is closely related to the number of available ALMA observations, and should not be used as the sole discriminator between variable and quiescent sources. In addition to returning variability statistics from ALMACAL data, after fitting, \textsc{RadioSED} produces plots like those shown in the bottom panels of Figure~\ref{fig:alma_variable}, where the best fitting SED model (as described below) is shown alongside data points coloured by observing epoch. The observing epoch is derived from the publication date of each survey in Table \ref{tab:survey_info}, except in the case of ACC data, where the observing epoch is explicitly provided as part of the catalogues. This allows the user to perform a simple, visual check as to whether the sources' apparent SED shape is at all influenced by the multi-epoch nature of the observations. However, for a more complete understanding of variability and in conjunction with the results from \textsc{RadioSED}, users may also find it useful to crossmatch against the Roma-BZCAT catalogue \citep{Massaro2015} to determine whether a source is a known blazar, as well as against the SPECFIND catalogue \citep{Stein2021TheFrequencies} to view flux densities spanning a broader time and frequency coverage. Furthermore, since $\gamma$-ray emitting AGN are almost always classified as blazars \citep{BoseGammaSources2022}, an additional cross-match with Fermi sources from, for example, the catalogue of \cite{DabruscoBlazars2019} may prove useful in identifying further blazar-like sources beyond those identified using radio SED variability alone. However, since these high-energy observations cannot be used in a radio SED, this functionality is not inbuilt in \textsc{RadioSED}, and our own study of the Fermi sources in our sample is left to future work.

Therefore, we do not expect variability to have a large effect on the SEDs constructed using the compiled surveys in Table~\ref{tab:survey_info} owing to the low rate and degree of variability amongst the radio AGN population. There may be some distortion of individual SEDs due to variability, but the largest impact is likely to be due to blazar contamination in any given peaked spectrum sample. However, since these flares appear transient over a period of months to years, our choice to compile flux densities from several surveys from different epochs in fact lowers the likelihood of such contamination, as does our use of the ALMA Calibrator Catalogue. In fact, \cite{Torniainen2007} note the advantage of compiling multi-epoch flux densities to isolate true peaked spectrum sources from larger samples. The number of sources we expect to be contaminating blazars or variable AGN is further discussed in Section \ref{sec:contamination}.

\section{Bayesian Modelling}\label{sec:bayesian_modelling}
With our SEDs constructed from the surveys listed in Table~\ref{tab:survey_info}, it remains now to fit them with suitable models in order to identify the peaked spectrum sources from amongst the larger population. To do this we employ a Bayesian approach, which allows us to not only infer the model parameters and their posterior probability densities for each model type, but also to compare different models and make an informed choice about which one best explains the data in question. Since our models can capture various degrees of curvature (as discussed below), we require a minimum of 4 flux density measurements at unique frequencies to perform our model inference. As we have been careful to construct our SEDs from large area surveys, many of which have flux limits on the order of a few milliJanskys, sources not meeting this minimum threshold are typically relatively faint in RACS-low ($S_{0.8\text{\,GHz}} \leq 15$\,mJy), though some may also be variables or transients.   

\begin{figure*}
    \includegraphics[width=\textwidth, trim={0cm 0.2cm 0 0},clip]{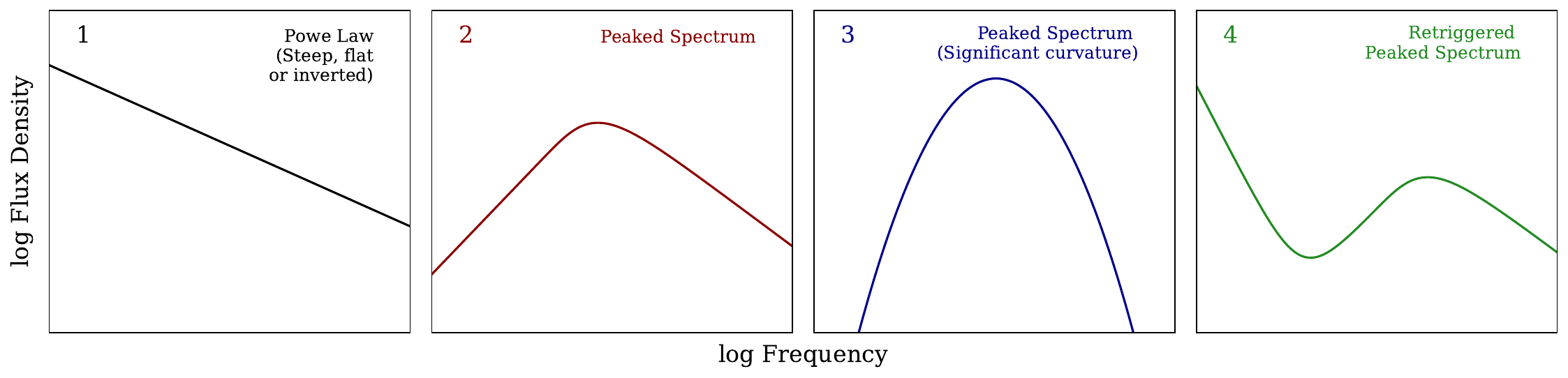}
    \caption{Indicative shapes of each of the 4 model types implemented by default as part of \textsc{RadioSED}.}
    \label{fig:model_examples}
\end{figure*}
\subsection{Models}\label{sec:model_types}

A set of four analytical models are applied to each SED using Bayesian inference. These are displayed in Figure \ref{fig:model_examples}, and each capture a physically distinct category of radio SED by modelling flux $S_\nu$ (or its logarithm) as a function of frequency $\nu$ as follows:\\ 

\textit{1) A steep, flat, or inverted spectrum source} 

\noindent Such sources are described by the simple functional form of a power law in Equation \ref{eqn:powlaw}. These three SED shapes (steep, flat, and inverted) vary functionally only by the sign of the spectral index $\alpha$, and hence are the simplest SEDs to fit. Physically, the flat spectrum sources are typically found to be quasars or non-flaring blazars, often with an elevated incidence of gamma-ray emission \citep[e.g.][]{Healey2007CRATES:Sources, Kimball2008, Mahony2010}, or the compact cores of extend radio sources which can be distinguished clearly on VLBI scales \citep[e.g.][]{Pushkarev2012}. The steep and inverted spectrum sources, by contrast, are thought to be peaked spectrum sources whose peak falls outside the observed frequency range, either due to being extremely compact (and hence exhibiting a high frequency peak), or extended out to $\sim10\,$kpc so that the peak frequency has shifted down into the megahertz regime \citep{ODea2021}. Their SED shape can be characterised by:

    \begin{equation}\label{eqn:powlaw}
        S(\nu) = S_0\nu^\alpha
    \end{equation}

where $S_0$ is the amplitude of the synchrotron spectrum in Janskys. By convention, the division between steep and flat-spectrum sources is often placed at a spectral index $\alpha = -0.5$ \citep[e.g][]{Urry1995} which we adopt here for consistency. Less well-defined is the distinction between flat and inverted spectrum sources, but we here adopt $\alpha = 0.5$ so that the range of flat spectrum indices is symmetric about zero.\\

\textit{2) A peaked spectrum source}

\noindent The simplest peaked spectrum source model is given in Equation \ref{eqn:snellen}. This was first used by \cite{Snellen1998} to characterise peaked spectrum sources, and it is now one of the more popular functional forms for describing these AGN, as it is simple to read off useful parameters like the peak frequency, peak flux and spectral indices \citep[e.g.][]{Callingham2017, Shao2020, Wolowska2021, Su2022}.

\begin{equation}\label{eqn:snellen}
    S_\nu = \frac{S_\text{p}}{1-e^{-1}}\left[1-e^{-(\nu/\nu_\text{p})^{\alpha_{\text{thin}} - \alpha_{\text{thick}}}}\right]\left(\frac{\nu}{\nu_\text{p}}\right)^{\alpha_{\text{thick}}}
\end{equation}

In the above equation, $S_\text{p}$ is the flux density at peak frequency $\nu_\text{p}$ and $\alpha_{\text{thin}}, \alpha_{\text{thick}}$ are the spectral indices in the optically thin (above $\nu_\text{p}$) and thick (below $\nu_\text{p}$) regions. This model reduces to the case of a homogeneous, synchrotron self-absorbed source when $\alpha_{\text{thick}} = 2.5$, but otherwise represents some inhomogeneous absorber which may be affected by either free-free absorption (FFA) or synchrotron self-absorption (SSA), such as a clumpy, circumnuclear medium \citep{Risalti2002}, or transient, radiation-driven outflows of ionised gas \citep{Wada2023}. Even when this function was first used to fit peaked spectrum sources, \cite{Snellen1998} noted that it sometimes failed to accurately capture SED structure far away from the peak, which brings us to our next two functional forms.\\

\textit{3) A peaked spectrum source with significant curvature}

\noindent Some peaked spectrum sources are not well fit by Equation \ref{eqn:snellen}. Instead, they can be described by a parabolic functional form in logarithmic space, or a log-parabola, as in Equation \ref{eqn:orienti}. 

\begin{equation}\label{eqn:orienti}
    \log S_\nu = a + \log \nu \left[b + c\log \nu \right]
\end{equation}

This form was originally used by \cite{Orienti2007} and \cite{Orienti2010} to characterise peaked spectrum sources because it was a good fit to their data, even though the parameters $a,b$ and $c$ were not thought to hold any physical meaning. However, \cite{Duffy2011TheSpectra} outline how the curvature parameter ($c$ here, $q$ in that work) can be used alongside the peak frequency $\nu_\text{p}$ and peak emissivity ($L_\text{p}$, found from the peak flux density $S_\text{p}$, distance, and size of the source) to derive the magnetic field strength and properties of the electron distribution in synchrotron emitting plasma lobes. \cite{Ross2021} used a similar log-parabola form to fit sources peaking at low frquencies in the GLEAM band, and the same function identified here is also used by \cite{Callingham2017} and \cite{Nyland2020} at radio frequencies. In the radio to sub-mm regime, a similar function is used to fit broadband, synchrotron self-absorbed emissions of blazars \citep[e.g.][]{Abdo2010TheBlazars, Chen2023}, where there is some evidence that the degree of curvature may correlated with peak frequency and is thus able to be related back to either statistical or stochastic particle acceleration under certain conditions \citep{Tracamere2011, Chen2014}.\\

\textit{4) A re-triggered peaked spectrum source}

\noindent In some cases, peaked spectrum sources exhibit an upturn again at the lowest frequencies, producing an SED well fit by the functional form in Equation \ref{eqn:retrig}. 

\begin{equation}\label{eqn:retrig}
    S_\nu = \frac{S_\text{p}}{1-e^{-1}}\left[1-e^{(\nu/\nu_\text{p})^{\alpha_{\text{thin}} - \alpha_{\text{thick}}}}\right]\left(\frac{\nu}{\nu_\text{p}}\right)^{\alpha_{\text{thick}}} + S_0\nu^\alpha
\end{equation}

This is a linear combination of Equations \ref{eqn:powlaw} and \ref{eqn:snellen}, with all of the same parameters from each of those. Physically, these sources are interpreted as AGN undergoing multiple cycles of activity on short timescales, or exhibiting distinct knots of emission along their jets, with the older, more diffuse emission producing the upturn at low frequencies \citep{Baum1990, EdwardsTingay2004, Hancock2010}. They are often, though not always, observed at the centre of cluster environments \citep{Hogan2015, Callingham2017}.

\subsection{Bayesian Inference}\label{sec:model_fit}

For each model, we test the hypothesis ($\mathcal{M}$) that it describes the true form of an SED given model parameters ($\pmb{\theta}$), and SED data (\pmb{$d$}), using Bayes' theorem:

\begin{equation}\label{eqn:bayes_theorem}
   \text{Pr}(\pmb{\theta} | \pmb{d}, \mathcal{M}) = \dfrac{\text{Pr}(\pmb{d}|\pmb{\theta}, \mathcal{M})\text{Pr}(\pmb{\theta} | \mathcal{M})}{\text{Pr}(\pmb{d}|\mathcal{M})}.
\end{equation}

The $\text{Pr}(\pmb{\theta}|\mathcal{M})$ term in the numerator of Equation \ref{eqn:bayes_theorem} is the prior probability, which encodes our belief about the parameters $\pmb{\theta}$ for model $\mathcal{M}$ \textit{a priori}. Since in our case we want to fit a large sample of sources about which we know very little, we want uninformative priors which make no assumption about the spectral shape of each source beyond constraining our parameter space to physically meaningful values. In general, this means we have used prior distributions uniform in logarithmic space (also known as a Jeffreys prior; \citealt{Jeffreys1946}) for scale parameters (i.e. peak frequency and flux), and distributions uniform in linear space for location parameters (i.e. spectral indices). The limits placed on these distributions are to ensure physically sensible parameters, meaning we restricted spectral indices to values $-4 < \alpha < 4$, peak flux values to $0.03 < S_{\text{peak}} < 15$\,Jy and peak frequency values to $0.05 < \nu_{\text{peak}} < 100$\,GHz across all models where these parameters are applicable. The only exception to this is the curvature parameter $c$ for the curved peaked spectrum model (Equation \ref{eqn:orienti}) for which we assume a truncated Gaussian prior which approximates the distribution of the curvature found for the sample described in \cite{Callingham2017} (there curvature is given as $q$). A more detailed breakdown of the priors for each model is provided in Table \ref{tab:priors}.

\begin{table}
	\centering
	\caption{Priors used in \textsc{RadioSED} for the four models outlined in Section \ref{sec:model_types}, plus priors for the Gaussian process used for SEDs with covariant GLEAM data as discussed in Section \ref{sec:implementation}. The \texttt{log\_const} factor in the Gaussian process represents the vertical scale of the kernel, and the M$_{00}$ factor represents the horizontal scale over which the kernel operates in frequency space.}
	\label{tab:priors}
	\begin{tabular}
{m{20mm}m{10mm}m{17mm}m{20mm}}
		\hline
		  Model & Parameter  & Prior Type & Prior range\\
		\hline
            \hline
		  Power Law & $S_0$ & log-uniform & 0.1 -- 10$\times10^7$\,mJy \\
		  (Equation \ref{eqn:powlaw}) & $\alpha$ & linear-uniform & -4 -- 4 \\ 
		\hline
             Simple peaked& $S_\text{p}$ & log-uniform & 30 -- 15$\times10^3$\,mJy \\
             spectrum & $\nu_\text{p}$ & log-uniform & 50 -- 10$\times10^4$\,MHz\\
             (Equation \ref{eqn:snellen}) & $\alpha_\text{thick}$ & linear-uniform & 0 -- 4\\
             & $\alpha_\text{thin}$ & linear-uniform & -4 -- 0\\
            \hline
            Curved peaked& $a$ & linear-uniform & -100 -- -1 \\
            spectrum & $b$ & linear-uniform & 1 -- 50\\
             (Equation \ref{eqn:orienti}) & $c$ & truncated Gaussian & $\mu=-0.5,\sigma=5$, [-1$\times10^{4}$, 0]\\
            \hline
            & $S_\text{p}$ & log-uniform & 30 -- 15$\times10^3$\,mJy \\
            & $\nu_\text{p}$ & log-uniform & 50 -- 10$\times10^4$\,MHz\\
            Re-triggered & $\alpha_\text{thick}$ & linear-uniform & 0 -- 4\\
            (Equation \ref{eqn:retrig}) & $\alpha_\text{thin}$ & linear-uniform & -4 -- 0\\
            & $S_0$ & log-uniform & 0.1 -- 10$\times10^7$\,mJy \\
            & $\alpha$ & linear-uniform & -4 -- 4\\
            \hline
            Gaussian process & \texttt{log\_const} & log-uniform & 3$\times10^{-7}$ -- 2 \\
            component & M$_{00}$ & log-uniform & 5 -- 30\,MHz \\
            \hline
	\end{tabular}
\end{table}

The $\text{Pr}(\pmb{d}|\pmb{\theta},\mathcal{M})$ term in Equation \ref{eqn:bayes_theorem} is the likelihood function $\mathcal{L}$, representing the probability of the SED data ($\pmb{d}$), given the current model $\mathcal{M}$ with parameters $\pmb{\theta}$. This is the key term that guides our model inference, as it allows us to update our belief about the model parameters using the available data. Since the data are assumed to be Gaussian distributed about the mean of model $\mathcal{M}$ with standard deviation given by the assumed noise of the signal(s), the probability density function in this case is a product of Gaussian PDFs over the vector of data. This is calculated simply using a multivariate Gaussian:

\begin{equation}\label{eqn:cov_likelihood}
    \mathcal{L} \equiv \dfrac{1}{\sqrt{(2\pi)^N} \det \mathbf{C}} \exp \left[-\dfrac{(\mathbf{d - m})^{\text{T}}\mathbf{C}^{-1}(\mathbf{d - m})}{2}\right]
\end{equation}

Here, \textit{N} is the length of the data vector $\mathbf{d}$, $\mathbf{C}$ is the covariance matrix for these observations, and $\mathbf{m}$ is the vector of modelled flux densities. As the GLEAM sub-band fluxes are known to exhibit some covariance \citep[][Section 5.4]{Hurley-Walker2017}, this full matrix formulation of the likelihood must be used for fitting sources with GLEAM data (and indeed, could be used for any sources with additional covariant data points introduced in the future). However, in cases where all data points are independent and uncorrelated, Equation \ref{eqn:cov_likelihood} reduces to:

\begin{equation}\label{eqn:likelihood}
    \mathcal{L} = \dfrac{1}{\sqrt{(2\pi)^N} \prod_i\sigma_i^2} \exp \left[-\sum_i\dfrac{(d_i - m_i)^2}{2\sigma_i^2}\right]
\end{equation}

\noindent where $\sigma_i$ is the uncertainty reported on flux density $\mathit{d_i}$. The posterior distribution of parameters $\pmb{\theta}$ for each individual model is inferred by traversing the prior parameter space until a stopping condition is reached where the model $\mathcal{M}(\overline{\theta})$ well describes the data with best-fitting parameter set $\overline{\theta}$. The exact implementation of this it outlined in Section \ref{sec:implementation} below.

\subsubsection{Censored data}

Since \textsc{RadioSED} is designed around several untargeted, large-area surveys, we can also make use of non-detections in a systematic way to better constrain sparse SEDs. For a source $t$ that falls within a survey area but does not have a reported flux density, we can assume an upper limit of $\overline{x}$\,Jy using the detection threshold given in the relevant survey paper. Therefore, a source may have flux density measurements $[d_1,...d_i]$ and upper limits $[\bar{d_{i+1}},..,\bar{d_k}]$. In such cases, we separate the likelihood into different components for observations and censored data. Following, for example, \cite{Klein2003CensoringTruncation}, we define the likelihood for the censored data using the cumulative distribution function of the likelihood for uncensored observations. In this case with a Gaussian likelihood for observations, for the censored data we will use the the error function (erf()) to calculate $Pr(d < \bar{d_k})$. In this way the likelihood function becomes:

\begin{equation}\label{eqn:censored_likelihood}
    \mathcal{L} = \prod_{x=1}^i\dfrac{1}{\sqrt{2\pi\sigma_x}}\exp{\dfrac{(d_x-m_x)^2}{2\sigma_x^2}}\prod_{y=i+1}^k \left[1 + \text{erf}\left(\dfrac{d_y-m_y}{\sigma_y}\right)\right]
\end{equation}

where $m$ is the flux density given by the model, $d$ is the observed/censored flux density, and $\sigma$ is the measurement uncertainty, as in Equations \ref{eqn:cov_likelihood} and \ref{eqn:likelihood}. All further steps proceed as with an SED comprising observed data only.

\subsection{Parameter Estimation}\label{sec:param_estimation}

\begin{figure*}
    \includegraphics[width=\textwidth, trim={0.5cm 1cm 0 0},clip]{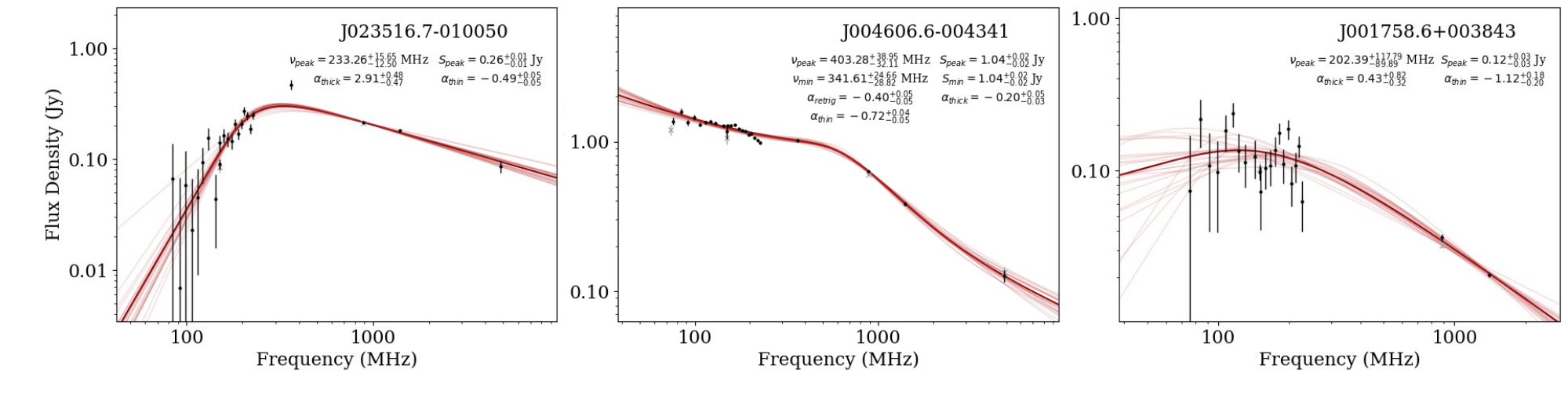}
    \caption{An example of each type of source which is best fit by a peaked spectrum-like model, but which requires some simple analysis of the fit parameters to fully understand the subtleties of the SED. Left is a source that would canonically be deemed flat spectrum with $\alpha_\text{thin} > -0.5$, but is classified here as a ``soft peaked spectrum'' source. Middle is a source which is best fit by the re-triggered peaked spectrum model, and yet all three spectral indices have $\alpha < 0$, indicating it is in fact a steep spectrum source with some degree of spectral curvature but no turnover. Right is a source best fit by Model~\ref{eqn:snellen}, and yet the fit parameters are poorly constrained with $\Delta{S_\text{peak}} \geq 0.4\times{S_\text{peak}}$, so the source is re-classified as complex. The parameter estimates with derived uncertainties for the best fitting model are shown in the top right of each sub-plot for reference. The dark red line is the best fit from the posterior of the selected model type, light red lines are random draws from the posterior to represent graphically the range of best-fitting parameters.}
    \label{fig:postprocessing}
\end{figure*}

Once the stopping condition is reached, we infer the most likely values for model parameters $\pmb{\theta}$ which can be further used for a more detailed and physically-motivated analysis the radio source in question. Using a random sample of draws from the posterior $\text{Pr}(\pmb{\theta} | \mathbf{d}, \mathcal{M})$ to create several realisations of model $\mathcal{M}$ , we derive a median value and 68\% credible interval on each parameter $\theta$.
In the case of models \ref{eqn:powlaw} and \ref{eqn:snellen}, we can directly obtain measures of spectral indices, and in Model \ref{eqn:snellen} we can also obtainthe peak frequency and flux. For models \ref{eqn:orienti} and \ref{eqn:retrig} inferring useful measurements like spectral indices is more complex as these are not directly encoded in the functional parameters. Instead for Equation \ref{eqn:orienti}, the peak frequency is derived as the axis of symmetry for a log-parabola: $\log(\nu_{\text{peak}}) = \frac{-b}{2c}$, with the corresponding peak flux $S_\nu$ obtained by evaluating the function at this frequency $\nu_\text{peak}$. Spectral indices $\alpha_\text{low}$ and $\alpha_\text{high}$ are derived by fitting a power law as in Equation~\ref{eqn:powlaw} either side of the peak, following \citep{Torniainen2005LongCandidates} and \citep{Orienti2007}. In the case of Equation \ref{eqn:retrig}, intervals for the peak frequency, flux and optically-thin spectral index, $\alpha_\text{thin}$ can be read from the function and returned, but we choose to derive these empirically from the data by identifying turning points in the curve, in addition to returning intervals for a pair of parameters we term the \textit{trough frequency} and associated \textit{trough flux}, where the SED reaches a local minimum before increasing again at lower frequencies. These aid in our identification of steep spectrum sources with mild curvature, which are sometimes well-described by this model and falsely classified as re-triggered peaked spectrum sources (more on this below).

\subsection{Model Comparison \& Selection}\label{sec:model_compare}

Once all four models given in Section \ref{sec:model_fit} have been applied to a single SED, our Bayesian inference method also allows us to robustly select which one of these is the preferred class of model thanks to the remaining term in the denominator of Equation \ref{eqn:bayes_theorem}.

This $\text{Pr}(\mathbf{d}|\mathcal{M})$ term is the Bayesian evidence $\mathcal{Z}$, which serves to normalise the posterior probability, and can be calculated numerically by marginalising the likelihood and prior of a model over that model's parameters $\pmb{\theta}$. This marginalisation means the evidence captures information about how well the overall model class $\mathcal{M}$ fits the data, bearing in mind all possible parameter values $\pmb{\theta}$ are explored from the prior distribution, not just the best fitting parameter set $\overline{\theta}$. In this way, a direct comparison of the evidence for different models enables us to determine which model $\mathcal{M}$ best fits the data without over-fitting and accounting for varying model complexity
\citep[e.g][Section 4]{Sivia2006}. This comparison term is usually called the Bayes factor $B_{1,2}$ and is found by taking the ratio:

\begin{equation}\label{eqn:bayes_factor}
    B_{1,2} = \dfrac{\mathcal{Z}_1}{\mathcal{Z}_2}
\end{equation}

so that a value of 2 means model $\mathcal{M}_1$ is twice as likely to be the true model as $\mathcal{M}_2$. In theory, the model with the highest Bayes factor from a pairwise comparison of all four models is will be the best fitting model without over-fitting the data.

In practice, there is always an uncertainty associated with the numerically-derived model evidence $\mathcal{Z}$ \citep[e.g.][Appendix A6]{Speagle2020}, and our choice of prior distributions (especially their limits) may have a mild effect on the calculated evidence, and hence on the derived Bayes factors. Accordingly, we follow the evidence thresholds of \cite{KassRafferty1995}, and consider any Bayes factors $0 < B_{1,2} < 0.5$ to represent only a marginal preference for $\mathcal{M}_1$ over $\mathcal{M}_2$. In instances where this is the case, and $\mathcal{M}_1$ is more complex than $\mathcal{M}_2$ (i.e. the number of parameters $N(\pmb{\theta_1})$ > $N(\pmb{\theta_2})$), we reject $\mathcal{M}_1$ as the most likely model to represent the data in favour of $\mathcal{M}_2$, following the principle of Occam's Razor.

\subsection{Source Classification}\label{sec:classify}

Once the most likely model type has been determined from a Bayesian analysis, the classification of the source is taken from this model following the definitions in Section~\ref{sec:model_types}. However, there are some subtleties in the use of these models and how they relate to the canonical definition of a peaked spectrum source, which should be taken into account when creating a robust sample of peaked spectrum AGN. We outline here three refinements to classifications that are performed as part of \textsc{RadioSED}, although we note that further refinement may be useful to the user after close inspection of their sources.

Firstly here, we note that there have historically been two main methods for identifying peaked spectrum sources. In small samples, often involving targeted followup observations, a peaked spectrum source has been defined as having an optically thin spectral index, $\alpha_\text{thin} < -0.5$ (where $\alpha < 0$ indicates a flux density which is decreasing as frequency increases), mimicking the canonical boundary between steep and flat spectrum sources \citep{Urry1995, ODea1998}, and a spectral curvature ($\alpha_\text{thick} + \alpha_\text{thin} > 0.6$) \citep{DeVries1997, EdwardsTingay2004}.

Across larger samples of up to a few thousand sources but with sparse flux density measurements, the identification has been done with as few as three independent flux densities which are used to create 2, two-point spectral indices at high and low frequencies (what constitutes `high' and `low' is not fixed but dependent on the availability of the data in any given study). These spectral indices are plotted against each other in radio colour-colour space, where one quadrant of the resultant plot with $\alpha_1 < 0$ and $\alpha_2 > 0$ indicates the presence of a spectral peak in some range of frequencies spanned by the flux density measurements used in calculation. This is the approach taken by \cite{Murphy2010TheCatalogue}, \cite{Callingham2017} and others (see e.g. Figure 15 in \cite{Murphy2010TheCatalogue} for a good example of this method), but it it does not provide a direct correspondence to the first method of identification. Indeed, \cite{Callingham2017} recognised the $\alpha_\text{thin} < -0.5$ requirement in the first method as an arbitrary threshold in a continuous spectrum of possible spectral indices, and instead used the radio colour-colour diagram to classify any sources with $\alpha_\text{low} > 0.1$ and $\alpha_\text{high} < 0$ as peaked spectrum for their work (note again that they use spectral indices $\alpha_\text{low}$ and $\alpha_\text{high}$ between fixed frequencies, rather than  $\alpha_\text{thin}$ and  $\alpha_\text{thick}$ determined by the location of the spectral peak). We strike a balance between these two approaches since our prior space for the peaked spectrum models extends beyond the $\alpha_\text{thin} < -0.5$ threshold (see Table \ref{tab:priors}), and we classify any source best fit by Models~\ref{eqn:snellen}, \ref{eqn:orienti} \& \ref{eqn:retrig} but with $0.1 > \alpha_\text{thin} > -0.5$ or $\alpha_\text{thick} + \alpha_\text{thin} < 0.6 $ a `\textit{soft peaked spectrum}' source. This allows for easy comparison with older, literature-derived samples, whilst also recognising the continuous distribution of spectral indices in physical sources.

Secondly, as mentioned in Section~\ref{sec:param_estimation}, some sources for which the most likely model is a re-triggered model (Model~\ref{eqn:retrig}), may in fact only exhibit slight spectral curvature and no discernible peak or turnover. In this case, the peaked classification is rejected in favour of a `complex' or steep/flat/inverted spectrum classification (dependent on the spectral indices inferred for the model).

At present, one final post-processing step is performed to identify poorly-constrained peaked spectrum classifications based on the parameter estimation described in Section \ref{sec:param_estimation}. If a source is classified as peaked spectrum, but the uncertainty on its peak frequency and/or peak flux density is so large that $\Delta\text{val} \geq 0.4\times\text{val}$, the peaked spectrum classification is marked as uncertain, but we do not modify the choice of most likely model. This step was included after a manual inspection of the synthetic data in Section \ref{sec:contamination} and the test sources discussed more in Section \ref{sec:stripe82}, and is found to identify a number of uncertain peaked spectrum classifications, which may occur due either intrinsic variability distorting the SED, or low S/N flux density measurements with large uncertainties.

Sources requiring each type of additional analysis for robust classification are shown in Figure~\ref{fig:postprocessing}. These are real sources drawn from our pilot field which will be discussed in more detail in the second paper in this series.

\subsection{Implementation}\label{sec:implementation}

\begin{figure}
    \centering
    \begin{tikzpicture}[node distance=2cm]
    \node (src_name) [startstop] {source name\\or position};
    \node (construct_sed) [process, below of=src_name, yshift=-0.5cm] {construct SED from public surveys\\(Table~\ref{tab:survey_info}, \\Sections~\ref{sec:included_surveys} \& \ref{sec:crossmatching})};
    \node (custom_sed) [startstop, right of=construct_sed, xshift=2cm] {custom SED data};
    \node (compactness) [io, below of=construct_sed, yshift=-0.5cm, xshift=2cm] {check source\\compactness \& flag \\ (Section~\ref{sec:compactness})};
    \node (variability) [io, below of=compactness, yshift=-0.5cm] {check source\\variability \& flag\\(Section~\ref{sec:variability})};
    \node (fit) [process, below of=variability, yshift=-0.5cm] {infer model parameters\\(Sections~\ref{sec:model_types} \& \ref{sec:model_fit})};
    \node (compare) [process, below of=fit, yshift=-0.5cm] {perform model\\comparison \& determine SED\\classification\\(Sections~\ref{sec:model_compare} \& \ref{sec:classify})};
    \node (output) [startstop, below of=compare, yshift=-0.5cm] {\textsc{RadioSED} output\\for further analysis};

    \draw [arrow] (src_name) -- (construct_sed);
    \draw [arrow] (construct_sed) -- (compactness);
    \draw [arrow] (custom_sed) -- (compactness);
    \draw [arrow] (compactness) -- (variability);
    \draw [arrow] (variability) -- (fit);
    \draw [arrow] (fit) -- (compare);
    \draw [arrow] (compare) -- (output);

    \end{tikzpicture}
    \caption{An overview of the \textsc{RadioSED} inference procedure for an individual source. Users can either input a source name/position and allow \textsc{RadioSED} to construct the SED using the method outlined in this paper, or input their own radio SED data to be fit.}
    \label{fig:radiosed_chart}
\end{figure}
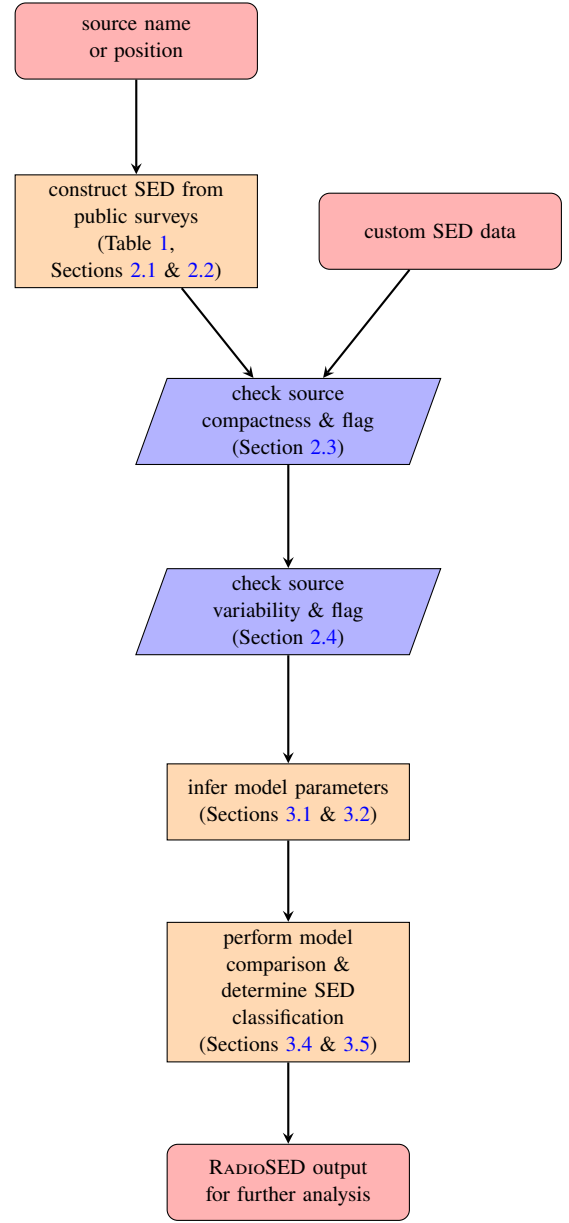

Our model inference is implemented in Python using the \textsc{Bilby} Bayesian Inference package \citep{Ashton2019Bilby:Astronomy} as an interface. \textsc{Dynesty} \citep{Speagle2020} performs the actual model inference using nested sampling, as this allows for robust evidence estimates which are necessary for model comparison \citep{SkillingNS2004, SkillingNS2006}. Within \textsc{Dynesty}, the dynamic nested sampling algorithm is used to improve performance by allocating the number of live points in a run dynamically, adding more points in a series of groups or \textit{batches} as the sampler moves towards successively higher areas of likelihood \citep{HigsondyNS19}. This allows the sampler to more accurately capture a complex posterior. The default behaviour for \textsc{RadioSED} is to ensure 1,000 effective samples in the final posterior or to progress through 10 batches, whichever comes first. For sources with GLEAM data, we also make use of the \textsc{George} Gaussian Process package \citep{georgepaper} to jointly fit for the data covariance and the model mean SED, since the exact form of this covariance is unknown \citep[][Section 5.4]{Hurley-Walker2017}.

As of publication date, \textsc{RadioSED} makes use of \textsc{Dynesty} version 2.1.2 \citep{dynestycode}, \textsc{Bilby} version 2.1.1, and \textsc{George} version 0.4.0. An overview of the steps involved in running \textsc{RadioSED} on a given source is shown in Figure~\ref{fig:radiosed_chart}.

\section{RadioSED Verification \& First Results}\label{sec:radiosed_verification}

\begin{figure*}
    \centering
    \includegraphics[width=\textwidth, trim={0 8.8cm 0 0},clip]{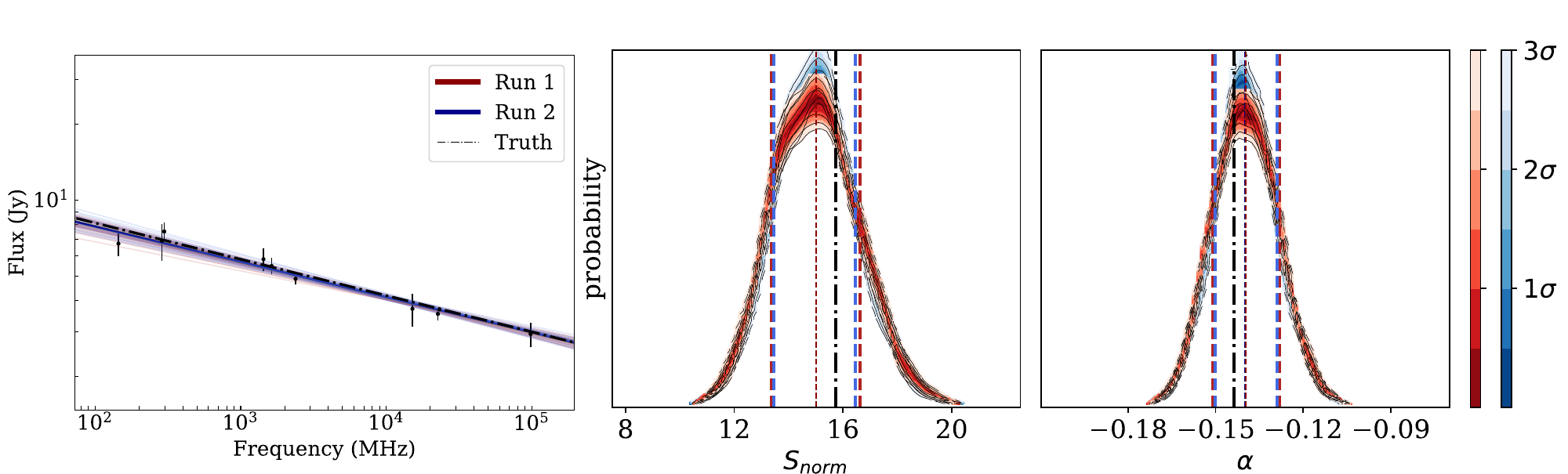}
    \caption{Left: output fits from \textsc{RadioSED} applied twice to a synthetic source obeying a power law of the kind in Equation~\ref{eqn:powlaw}. Dark lines show the best fit from the model posterior for each run (red and blue for runs 1 and 2, respectively), and lighter lines show random draws from the posterior. Dot-dashed line in black is the true SED of the source, from which observations are drawn with a 10\% Gaussian scatter.Right: diagnostic plots showing the pdf as a function of model parameters and produced by \textsc{Nestcheck} for repeated fitting of this same source. Dot-dashed black lines indicate injected parameter values, dotted coloured lines indicate output parameter means and dashed lines are the associated 1-$\sigma$ errors.}
    \label{fig:nestcheck_powlaw}
\end{figure*}
\begin{figure*}
    \centering
    \includegraphics[width=\textwidth, trim={1mm 2mm 0 0},clip]{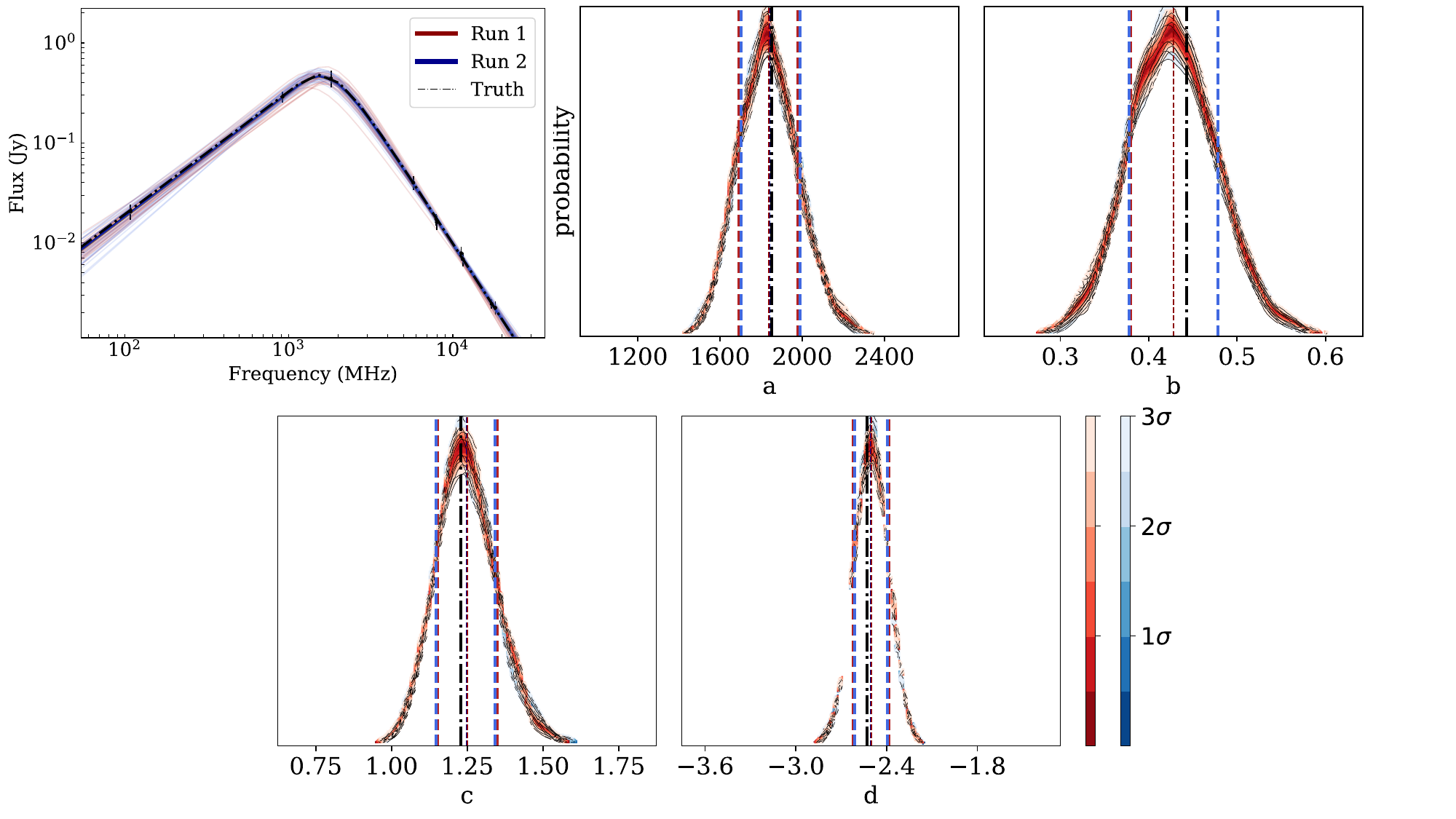}
    \caption{The same as Figure~\ref{fig:nestcheck_powlaw} but for a source obeying Model~\ref{eqn:snellen}.}
    \label{fig:nestcheck_snellen}
\end{figure*}

We outline here a first application of \textsc{RadioSED} to both synthetic and real-world data, demonstrating its accuracy and reliability as a tool for identifying young, obscured AGN from radio continuum data alone. We focus first on a discussion of individual models and potential sources of uncertainty in the modelling procedure, before moving on to consider whether our inhomogeneous SEDs could lead to contamination of any eventual peaked spectrum sample in the context of AGN variability. Finally, we examine the performance of \textsc{RadioSED} in recovering the SEDs of radio sources with well-characterised spectral shapes from the literature.

\subsection{Parameter Recovery \& Sampling Errors: Nestcheck }\label{sec:nestcheck}

Here we examine the reproducibility of the results output by \textsc{RadioSED} using synthetic broadband flux density measurements. We validate output for each of the 4 models outlined in Section \ref{sec:model_types} separately as detailed below.

\begin{figure*}
    \centering
    \includegraphics[width=\textwidth, trim={1mm 2mm 0 0},clip]{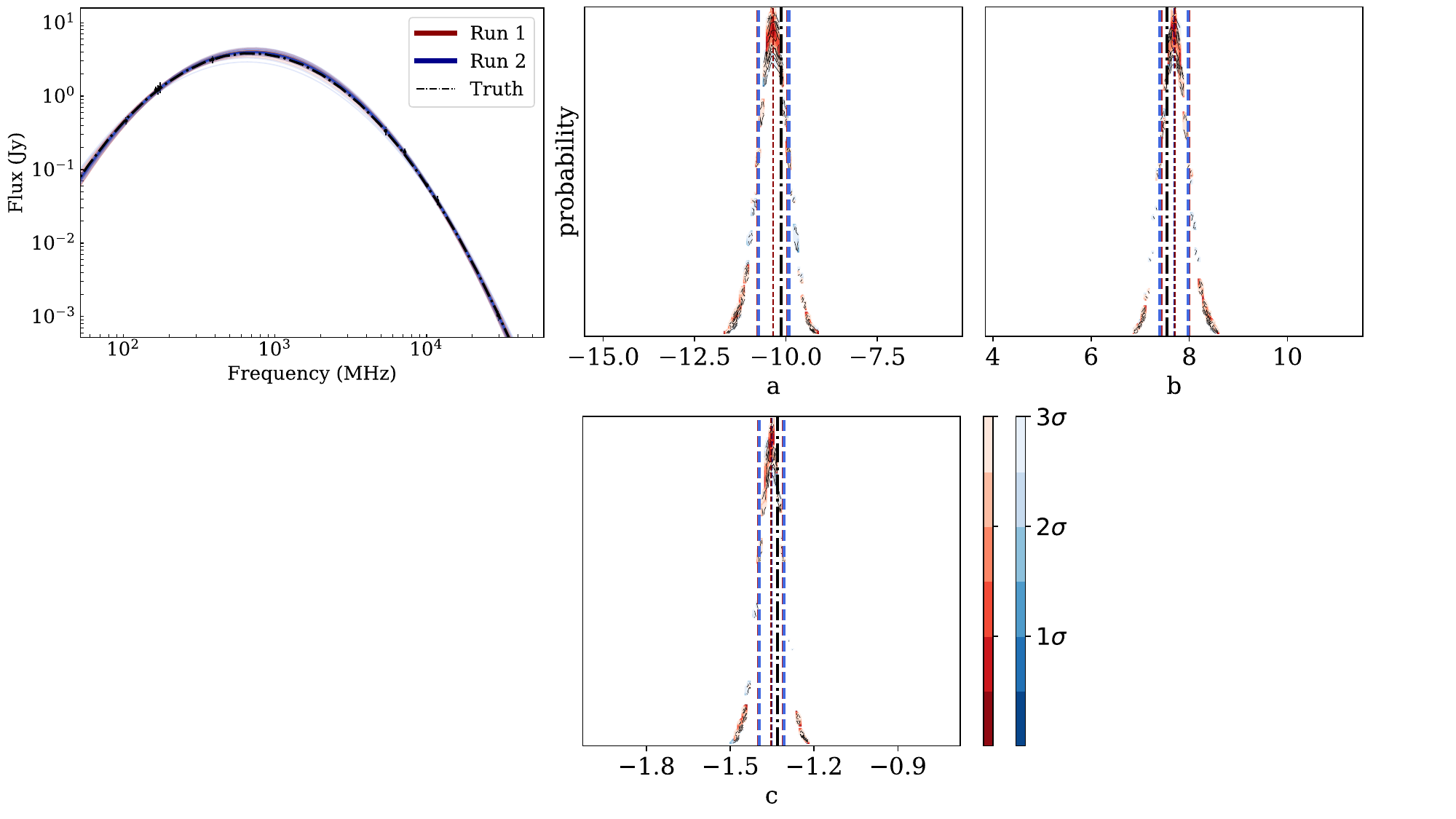}
    \caption{The same as Figure~\ref{fig:nestcheck_powlaw} but for a source obeying Model~\ref{eqn:orienti}.}
    \label{fig:nestcheck_orienti}
\end{figure*}
\begin{figure*}
    \centering
    \includegraphics[width=\textwidth, trim={1mm 30mm 0 0},clip]{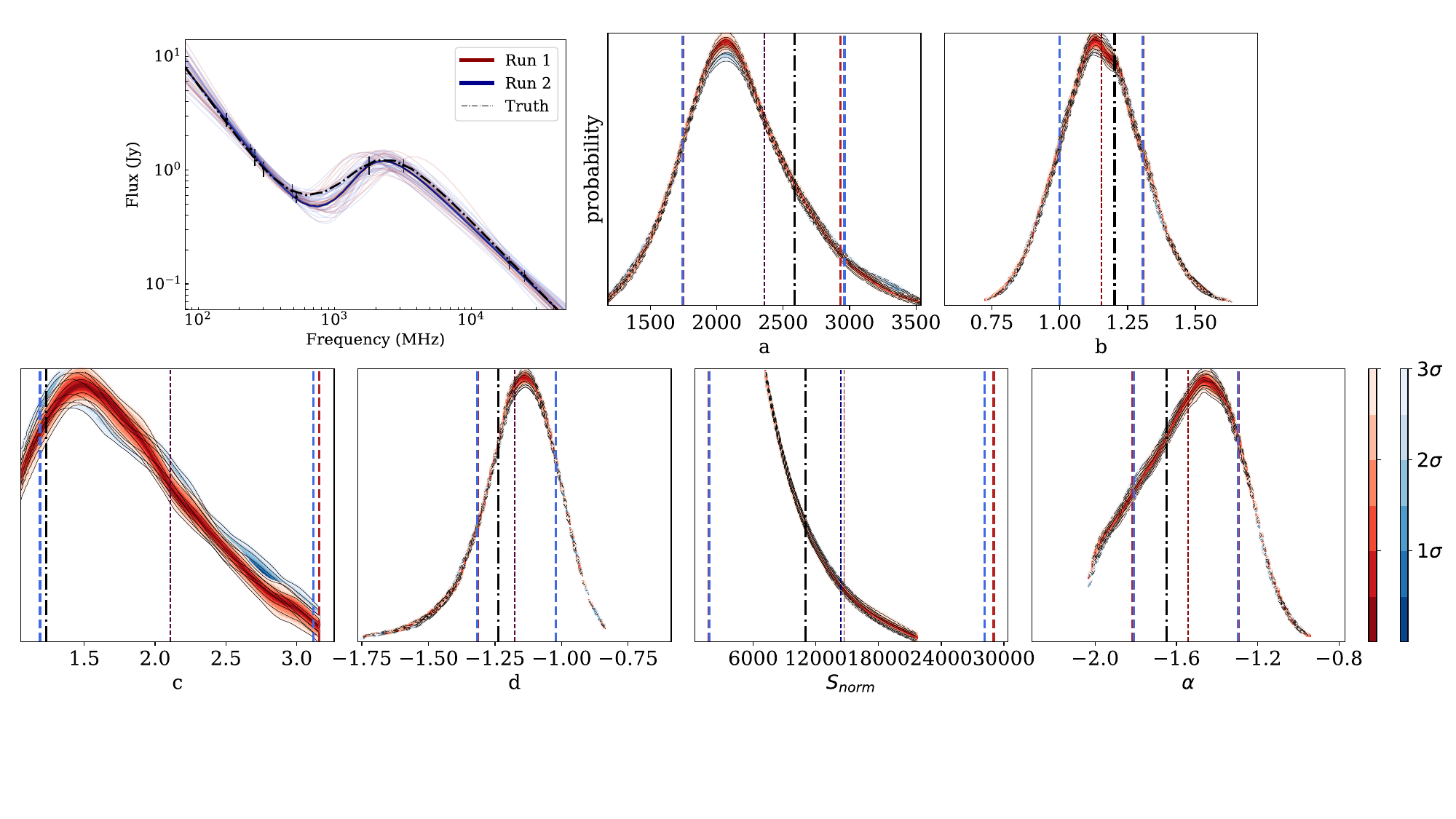}
    \caption{The same as Figure~\ref{fig:nestcheck_powlaw} but for a source obeying Model~\ref{eqn:retrig}.}
    \label{fig:nestcheck_retrig}
\end{figure*}

Since nested sampling by construction never reaches a steady state, but proceeds always towards regions of the posterior with a higher likelihood, it can be difficult to validate nested sampling-based approaches. In particular, quantifying implementation-specific uncertainties is tricky as we cannot rely on autocorrelation-based heuristics which are typically used in MCM approaches to determine whether successive runs will produce reproducible output (i.e. whether the runs converge and the posterior samples are suitably uncorrelated). Therefore, to ensure that the fits produced by \textsc{RadioSED} are accurate and reproducible, we make use of the \textsc{Nestcheck} package \citep{HigsonNestcheck2019}, which is designed explicitly to address this difficulty. The package provides three main useful features; first it characterises the uncertainty of a nested-sampling algorithm between successive runs, secondly it quantifies implementation-specific sources of error, and finally it provides useful diagnostic plots to summarise these effects. This third feature, the diagnostic plots, also means the output of \textsc{Nestcheck} can be easily adapted to test the ability of \textsc{RadioSED} to recover injected parameters as well.
\begin{figure*}
    \includegraphics[width=\textwidth, trim={0 0cm 0 0},clip]{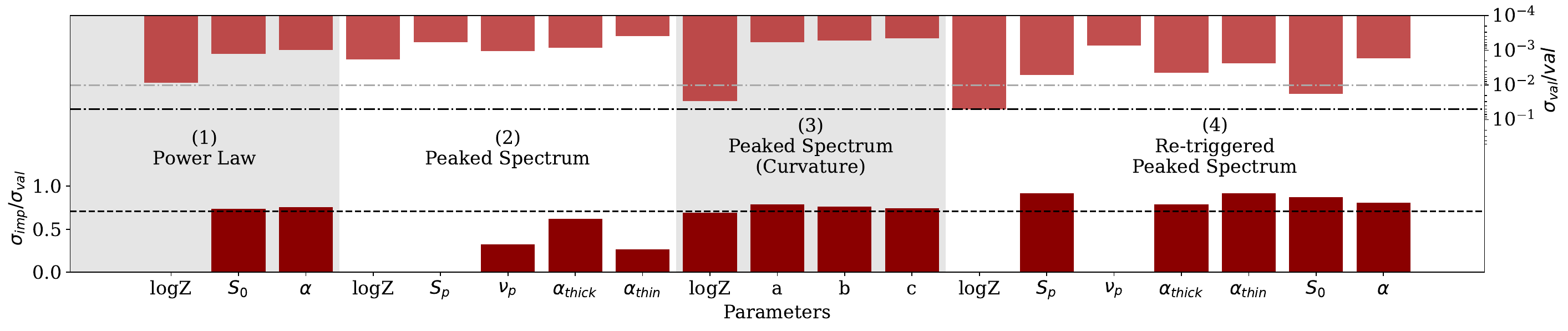}
    \caption{Summary plot showing the small fractional uncertainties in model parameters derived using \textsc{RadioSED} (top), as well as breaking down the uncertainty into contributions from implementation effects, and from the inherent stochasticity of nested sampling algorithms (bottom). All uncertainties sit close to or below the black dot-dashed (top) and black dotted (bottom) lines, indicating that \textsc{RadioSED} will be sufficiently accurate for deriving SED parameters with which to perform meaningful, physics-driven analysis. Uncertainties were derived using the \textsc{Nestcheck} package. The black dot-dashed line indicates a fractional error of 5\%, and the grey a fractional error of 1\%.}
    
    \label{fig:nestcheck_errs}
\end{figure*}

We consider first potential inter-run variations of \textsc{RadioSED}. The nested sampling algorithm is stochastic in nature, and so some low-level variation of the posterior and evidence output between runs is to be expected. However, if this variation is too high, estimates of model parameters and Bayesian evidence become unreliable. We do not expect a large inter-run variation here as \textsc{RadioSED} makes use of a standard and robust nested sampling package which incorporates stopping criteria based on minimising uncertainty in both the posterior and evidence \citep[][Appendix A]{Speagle2020}. Indeed this is confirmed by \textsc{Nestcheck}'s diagnostic plots, shown in Figures \ref{fig:nestcheck_powlaw} - \ref{fig:nestcheck_retrig}. These plots present the uncertainty of the posterior of two separate nested sampling runs for a synthetic spectrum representing each model type outlined in Section \ref{sec:model_fit}. The synthetic spectra were generated by sampling from the prior for each model type to create an SED shape, and then sampling from this SED at 10 random frequencies to create flux density measurements. To emulate real data for a non-varying source, these flux densities were given a 5\% Gaussian scatter, with uncertainties set at 10\% of this flux value, plus an additional component of Gaussian noise, again at the 5\% level. After running \textsc{RadioSED} twice, uncertainties on each posterior were obtained by bootstrap re-sampling the individual runs \citep{Higson2018}. The fact that both posterior probability  distributions, indicated by the different colours, occupy similar regions of probability space demonstrates that the stochasticity of nested sampling does not have a substantial effect on the estimates output by \textsc{RadioSED}.

Secondly, and perhaps more importantly, we consider implementation-specific effects. These may arise due to everything from subtle inaccuracies in the construction of the prior and likelihood distributions, to the structure of the nested sampling software itself \citep{HigsonNestcheck2019}. It is therefore possible that such effects may arise in \textsc{RadioSED} due to elements of the code which build upon established packages. \textsc{Nestcheck} derives the uncertainty in the Bayesian evidence ($\log \mathcal{Z}$) and parameter estimates due to implementation-specific effects by calculating the uncertainty over several runs and comparing this to the uncertainty expected from stochasticity alone (calculated via bootstrapping a single run, as discussed above). These calculations were performed using 10 runs for each model type, and the results are summarised in Figure~\ref{fig:nestcheck_errs}.

In Figure~\ref{fig:nestcheck_errs}, the bars at the bottom indicate the fraction of the total standard deviation $\sigma_{\text{var}}$ which is due to implementation-specific effects. If this value exceeds $\frac{1}{\sqrt{2}}$ as indicated by the dashed horizontal line, it means that the uncertainty in a parameter estimate is due mainly to implementation effects, if it is less than this threshold then the uncertainty is largely caused by the inherent stochasticity of nested sampling. This is due to the expected form of the variance, as outlined in \cite{HigsonNestcheck2019}, where $\sigma_\text{val}^2 = \sigma_\text{imp}^2 + \sigma_\text{bootstrap}^2$.  For both the power law (Equation \ref{eqn:powlaw}) and simple peaked spectrum (Equation \ref{eqn:snellen}) models, the stochasticity of nested sampling dominates the uncertainties, while for both the peaked spectrum with curvature (Equation \ref{eqn:orienti}) and re-triggered models (Equation \ref{eqn:retrig}), implementation effects dominate the uncertainty of all except the Bayesian evidence ($\log \mathcal{Z}$). However, this is only meaningful when considered in the context of the overall fractional uncertainty on these values, which is shown along the top of Figure~\ref{fig:nestcheck_errs}. As we can see, the fractional error is below 5\% for all parameters and evidence values (the dot-dashed black line), and below 1\% (dot-dashed grey line) for all except the Bayesian evidence for models \ref{eqn:orienti} and \ref{eqn:retrig}, and the normalisation factor for model \ref{eqn:retrig}. Since the Bayesian evidence is used in model comparison and selection, the fact that the uncertainty on these values for models \ref{eqn:orienti} and \ref{eqn:retrig} is at the 5\% level reinforces our choice to implement a correction to the model selection procedure, as outlined in Section \ref{sec:model_compare}.
Overall though, while implementation-specific effects contribute equally or more than stochastic effects in models \ref{eqn:powlaw}, \ref{eqn:orienti} and \ref{eqn:retrig}, the fractional uncertainties remain sufficiently low for all parameters of interest. 

Because the diagnostic plots produced by \textsc{Nestcheck} show the uncertainty on the posterior distribution with reference to the median value of the model parameters returned by a sampling run, it is also possible to use these results to validate how well \textsc{RadioSED} can recover the true parameters for various SED shapes. In Figures \ref{fig:nestcheck_powlaw} - \ref{fig:nestcheck_retrig}, the dotted red and blue lines indicate the mean values for each parameter derived from each nested sampling run, while the dashed lines represent 68\% confidence intervals on these values as derived from the posterior distributions. To make these plots useful for verifying how well \textsc{RadioSED} can recover the true parameters of an SED, we have modified \textsc{Nestcheck}'s diagnostic diagrams, adding the true injected parameter value for each spectrum as the dot-dashed line in black. We have also added an additional sub-plot (the leftmost sub-plot in each Figure) showing the synthetic SED along with the best output from each run of \textsc{RadioSED}, 50 draws from the posterior to give an indication of parameter range, and the true SED shape. In all cases \textsc{RadioSED} has recovered the injected parameter values to within the 1-$\sigma$ uncertainties shown in the figures, demonstrating it is suitably accurate for use on real-world observations.

\subsection{Variability \& Sample Contamination}\label{sec:contamination}
As mentioned in Section \ref{sec:variability}, it is conceivable that the radio variability of AGN could lead to false detections of peaked spectrum sources when combining multi-epoch flux density measurements. While the use of the ALMA Calibrator Catalogue provides one measure of variability at sub-mm frequencies, variability across much of the frequency space used in our modelling is difficult to quantify from the surveys incorporated into \textsc{RadioSED}. Therefore, to better understand the rate of false peaked spectrum detections due to variability, we have undertaken some simple, Monte Carlo-based simulations. 

\begin{figure}
    \centering
    \includegraphics[width=0.48\textwidth]{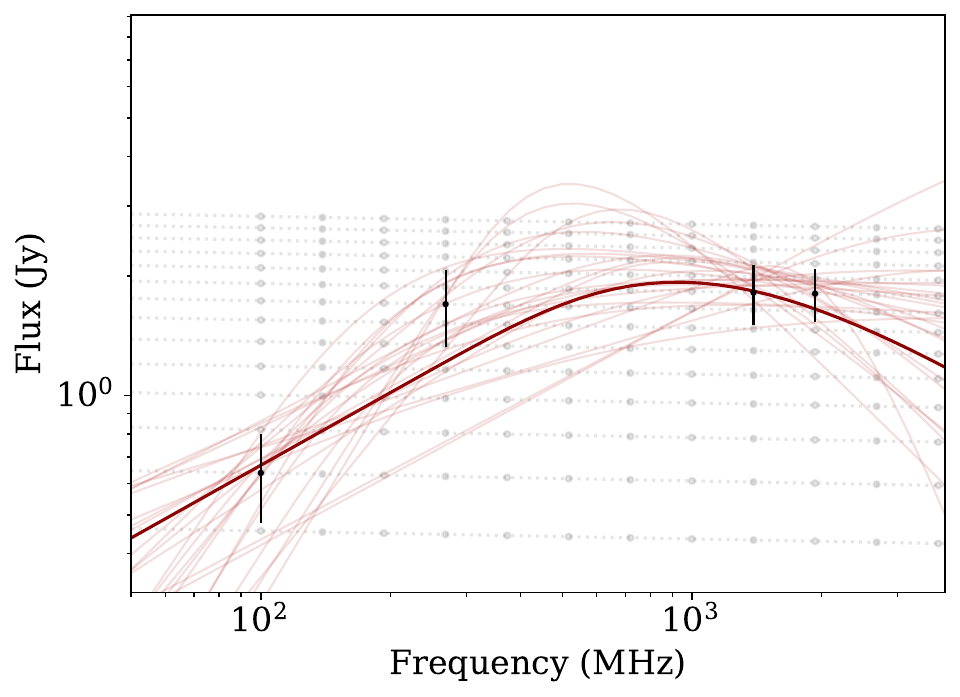}
    \caption{Mock variable flat spectrum source (dotted grey lines), with an overlaid ``multi-epoch'' SED (black points) and fit from \textsc{RadioSED} (dark red). Uncertainties are 10\% of the flux value plus an additional 10\% Gaussian scatter.}
    \label{fig:variable_master}
\end{figure}

We began by creating a mock-variable flat spectrum source from which to draw 'multi-epoch' observations. This synthetic source, shown in Figure~\ref{fig:variable_master}, was created by using Equation \ref{eqn:powlaw} with a spectral index $\alpha = -0.02$ and an amplitude varying between $0.5 \leq S_0 \leq 3.2$, to mimic broadband spectral variability. We chose a flat spectrum source since \cite{McConnell2012ATPMN:Survey} found sources with $\alpha > -0.5$ exhibited most variability between 4.8-8.65\,GHz. Overlaid on this variable source is a mock set of five multi-epoch observations, made by selecting a single flux density measurement at 5 unique frequencies. By repeatedly sampling this synthetic source using between 4-15 unique frequency `measurements', we can construct a suite of mock SEDs to be run through \textsc{RadioSED}, using a Monte Carlo method to explore false detection rates as a function of $N_{\text{obs}}$, the number of observations at unique frequencies.

\begin{figure}
    \centering
    \includegraphics[width=0.48\textwidth]{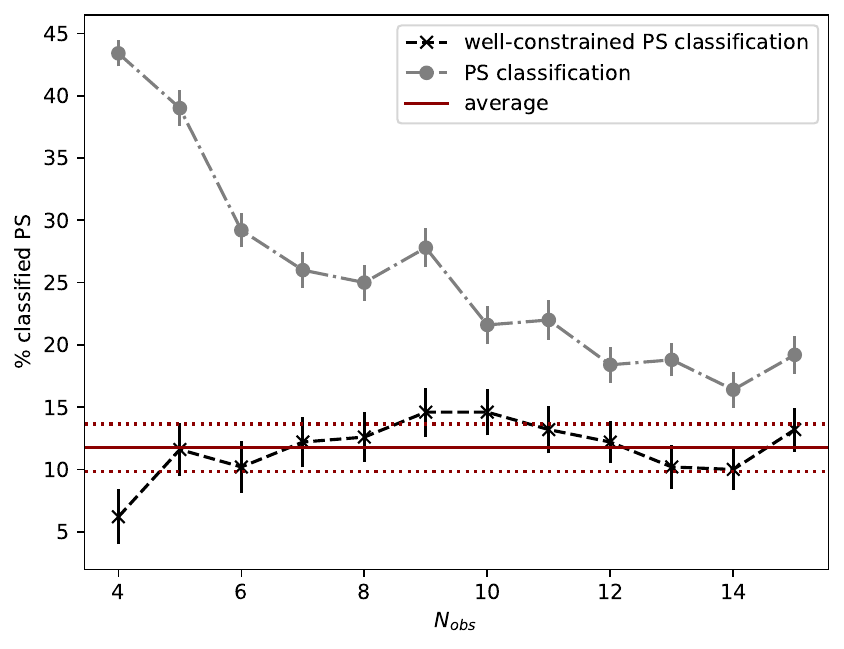}
    \caption{Fraction of mock variable SEDs misclassified as peaked spectrum due to multi-epoch sampling. Note especially that the rate appears much higher if we consider all sources misclassified as peaked (grey dot-dashed), and do not discount the unreliable or poorly constrained ones (black dashed) as defined in Section \ref{sec:contamination}. This shows the importance of our reliability criteria. Uncertainties on all measurements are bootstrapped. The dark red line shows the average misclassification rate across all $N_{\text{obs}}$, and the red dotted lines are the $1\sigma$ uncertainties on this value. }
    \label{fig:classification_frac}
\end{figure}

After creating 500 mock SEDs per unique $N_{\text{obs}}$ and running these through \textsc{RadioSED}, the false detection rate of peaked spectrum sources was found to be approximately 11.8$\pm1.9$\% from amongst variable, flat spectrum sources. This was after discounting sources unreliably classed as peaked spectrum, as defined in Section \ref{sec:contamination} and shown in Figure \ref{fig:classification_frac}. We also found that in 96\% of mis-classified sources, the false peak was three measurements or fewer from the edge of the observing band, which may be a helpful heuristic for identifying mis-classified sources in the future, though further work will need to be done in comparing this with real data for secure peaked spectrum sources.

Considering this in the broader context of peaked spectrum samples, we expect the overall contamination of any peaked spectrum sample produced using \textsc{RadioSED} to be low. If we assume conservatively that 5\% of radio AGN are significantly variable in some region of observed frequency space, based on the discussion in Section \ref{sec:variability}, and we expect 12\% of these to be mis-classified as peaked spectrum from our simulations, that is only 0.6\% of the radio population mis-classified as peaked spectrum. Or, put another way, if 15\% of an unbiased radio population is classified as peaked spectrum, we expect about 4\% of these classifications to be incorrect. This is much lower than the mis-classification fractions seen in for example \cite{Torniainen2007} and \cite{EdwardsTingay2004}, who re-observed known peaked spectrum sources identified in single epochs, and revealed variability in a large fraction of them by utilising a multi-epoch analysis. Since \textsc{RadioSED} begins from a multi-epoch approach, such contamination is in fact likely to be much lower in the peaked spectrum samples produced in the method outlined in this work. A deeper consideration of radio variability and its impact on multi-epoch radio SEDs will be considered as part of our second paper in the context of a larger sample of radio sources with real world, observational data.

\subsection{Test Sources}\label{sec:stripe82}
With the characteristics of \textsc{RadioSED} better understood we now apply it to a selection of radio sources with various published SED shapes. The purpose of this exercise is to determine how well our framework can recover expected spectral parameters for sources known to fall into each of the four categories given by equations \ref{eqn:powlaw} - \ref{eqn:retrig}.

\begin{figure*}
    \centering
    \includegraphics[width=\textwidth
    ]{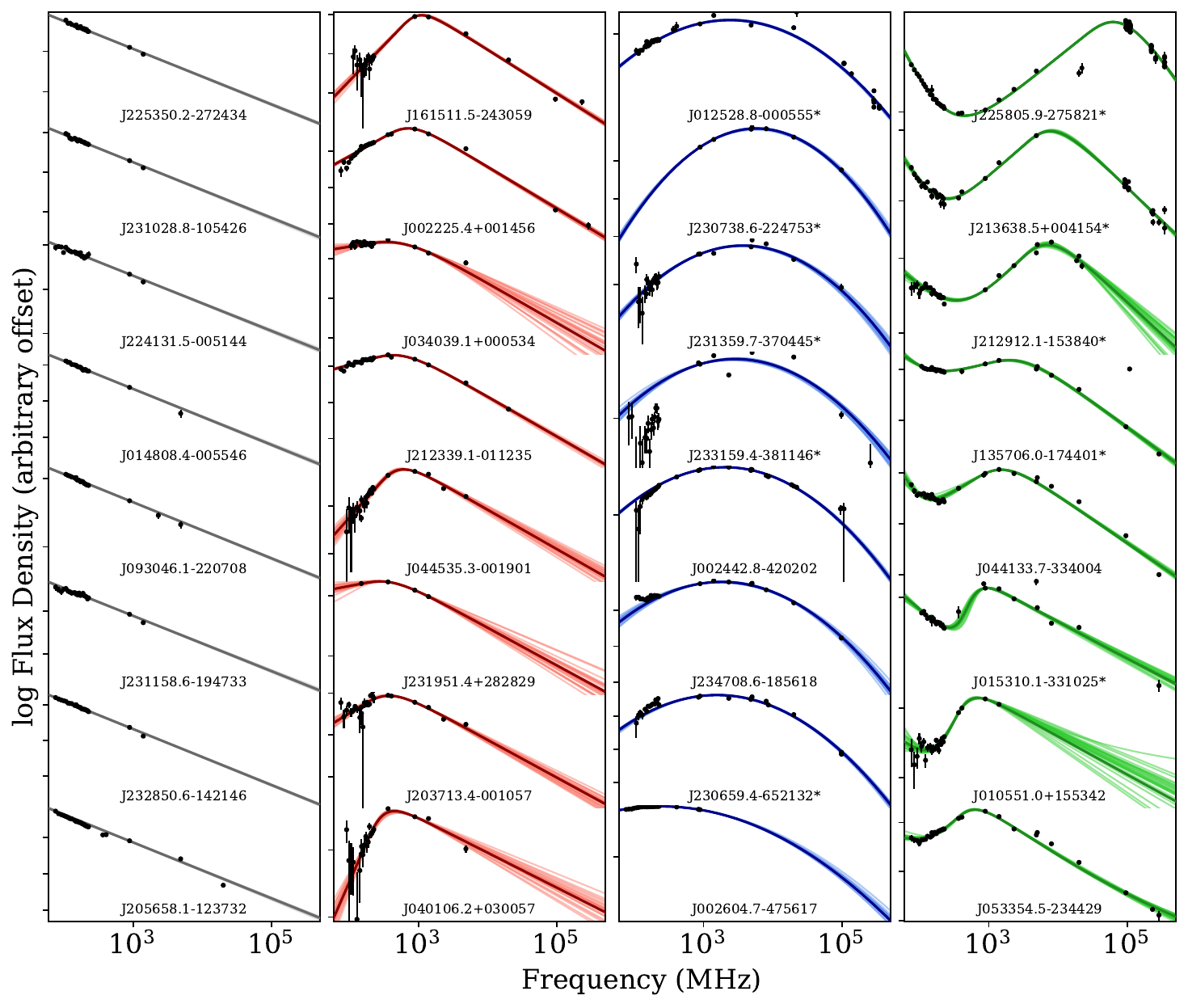}
    \caption{A representative sample of sources from the literature with well-characterised SED shapes used to verify that \textsc{RadioSED} is able to recover expected spectral parameters. Columns map onto the different SED shapes from Figure \ref{fig:model_examples}, so these sources have previously been identified as having a power law SED corresponding to Model \ref{eqn:powlaw} (left column), a peaked spectrum following Model \ref{eqn:snellen} (second left), a peaked spectrum with significant curvature after Model \ref{eqn:orienti} (right middle), or an upturn at lower frequencies (right), fit by Model \ref{eqn:retrig} as described in Section \ref{sec:model_types}, and shown in Figure \ref{fig:model_examples}. The y-axis for each individual source has been dynamically scaled to show the full extent of the model between 100\,MHz-500\,GHz, which causes the slopes of all the Power Law sources (left column) to appear almost identical. For each source, the darkest line shows the best fit of the most probable model from \textsc{RadioSED}, while the array of lighter lines represents 25 draws from the posterior of this model to give an indication of the spread in parameter values.}
    \label{fig:observed_sources}
\end{figure*}

To obtain a selection of power law (Model \ref{eqn:powlaw}) and peaked spectrum (Model \ref{eqn:snellen}) sources, we consulted the SPECFIND V3.0 catalogues \citep{Stein2021TheFrequencies}. We constrained our search to spectra with at least 8 flux density measurements in the final SPECFIND SED, and a declination $-30^{\circ} < \delta < 30^{\circ}$ to obtain sources within the area where the RACS-low and NVSS surveys overlap. In addition to SPECFIND, we consulted \cite{Callingham2017} to obtain a selection of sources with a low-frequency upturn (Model \ref{eqn:retrig}), as these sources were explicitly discussed in Section 6 of that work, and a manual inspection of the SEDs for sources in the AT20G high angular resolution catalogue \citep{Chhetri2013TheCatalogue} revealed a number of peaked spectrum sources with significant curvature based on their changing spectral indices as a function of flux density (Model \ref{eqn:orienti}).  Some additional sources from the literature known to the authors were further added to this sample, to allow us to present an equal number of sources within each model category. The fits of these sources obtained using \textsc{RadioSED} are given in Figure \ref{fig:observed_sources}, and the parameters of their fits are presented in Table \ref{tab:obs_sources}.

The parameters given by \textsc{RadioSED} are broadly consistent with those found in the literature for this sample of sources, bearing in mind the constraints under which many literature parameters were originally derived. For example, peaked spectrum sources identified in \cite{Stein2021TheFrequencies} assume a fixed peak frequency of either 325\,MHz or 1.4\,GHz, and the spectral indices are typically derived as $\alpha_{low}$ and $\alpha_{high}$ between fixed frequencies, rather than $\alpha_{\text{thick}}$ and $\alpha_{\text{thin}}$ either side of the peak, as is done in \textsc{RadioSED}. More importantly, it is apparent from Figure \ref{fig:observed_sources} that our framework can appropriately model SEDs spanning this diverse range of shapes and spectral parameters. The successful fitting of these sources is therefore a promising, early indication of the of the efficacy of \textsc{RadioSED} in large, statistical studies of radio spectral shapes.

However, as noted in Sections \ref{sec:variability} and \ref{sec:contamination}, there remains the possibility of blazar contamination amongst any given peaked spectrum sample. Indeed from amongst these sources, a search of the SIMBAD database \citep{Wegner2000} revealed 10 sources with either a primary or secondary blazar classification. The names of these sources are marked with an asterisk in Figure \ref{fig:observed_sources} and Table \ref{tab:obs_sources}. Of these, 9 sources have blazar classifications derived from the \cite{Massaro2015} Roma-BZCAT catalogue where they are classed as `quasar-like' blazars, meaning they were identified on the basis of a flat radio spectrum between 0.8-5\,GHz and broad, optical emission lines. The remaining source, J230738.6-224753, is identified as a blazar candidate in the CGRaBS survey \citep{healey2008}. However, the broadband SED modelling performed here reveals significant curvature in each of the blazar-identified `flat spectrum' sources, highlighting the need for broadband coverage when searching for beamed radio emissions. Indeed in the original Roma-BZCAT paper, \cite{Massaro2009} noted that GPS sources and blazars are sometimes confused in radio-selected samples due to poor spectral coverage ``both in frequency and time''. As a result of this difficulty, any known GPS sources otherwise meeting the Roma-BZCAT criteria for a blazar were included in the first edition of that catalogue. Of course in the case of a secure blazar classification, an object cannot be a true Peaked Spectrum source in the canonical sense of young radio AGN with minimal variability, but it is clear that we must be careful to check any such classification against the full set of mutliwavelength data available for a source. In the case of the blazar-identified sources shown here, many do exhibit high X-ray luminosities and $\gamma$-ray emissions consistent with true blazars, but we have still chosen to include their radio SED fits for demonstrative purposes. Not only do they reinforce the ability of \textsc{RadioSED} to recover peaks previously identified in the literature, but they also emphasise the need for additional tools, beyond SED analysis, to identify true Peaked Spectrum sources. We therefore recommend to any users of \textsc{RadioSED} that a crossmatch with up to date blazar catalogues be used in conjunction with the outputs from our framework, and indeed our forthcoming work on our chosen pilot field will make use of just such ancillary data.

\section{Conclusions \& Future Work}\label{sec:conclusions}

In this paper we have presented \textsc{RadioSED}, a new SED fitting framework tailored to broadband, radio frequency observations and capable in particular of identifying peaked spectrum sources across a wide range of ages and redshifts. \textsc{RadioSED} incorporates pre-matched, large area radio catalogues spanning $0.072-370$\,GHz to provide the user with meaningful, consistent SEDs with which to do science (Section \ref{sec:sed_construction}), and these SEDs are reliably fit using a rigorous Bayesian approach (Section \ref{sec:bayesian_modelling}). We have validated \textsc{RadioSED}'s ability to recover injected parameters by making use of the \textsc{Nestcheck} package (Section \ref{sec:nestcheck}), and our own simple Monte Carlo simulations have shown that the misclassification rate of variable sources in any given sample is likely to be low, sitting at $11.8\pm1.9\%$ (Section \ref{sec:contamination}). Not only that, but by applying this framework to a representative sample of sources with well-known SED shapes from the literature, we have shown that \textsc{RadioSED} can reliably recover the expected spectral properties of real radio sources using only the radio surveys outlined in Section \ref{sec:included_surveys}, and the models from Section \ref{sec:model_types}. 

As to future work, we are currently completing an investigation into the radio AGN within the Stripe 82 field (Kerrison et al. in prep), for which a full, multiwavelength analysis is likely to reveal important details about their host galaxies and local environments \citep[e.g.][]{Hogan2015}. A full investigation of this test field in comparison to synthetic spectra like those discussed here in Section \ref{sec:contamination} will also help to further clarify the effects of variability on multi-epoch SED fitting, all of which will pave the way for larger, more reliable, and more complete samples of radio AGN in various classes. Alongside this, the modular nature of \textsc{RadioSED} means that it can be continually improved as the next generation of large area surveys collect in public repositories, and our understanding of the nuances behind radio SED modelling grows. In particular, new and forthcoming IPS measurements at both megahertz and gigahertz frequencies could aid greatly in the identification of compact sources \citep{Morgan2022, Chhetri2023}, while the GLEAM-X (72-231\,MHz; \cite{HurleyWalkerGLEAMX}) RACS-mid (1.36\,GHz; \cite{Duchesne2023}) and RACS-high (1.65\,GHz) surveys will provide additional continuum data to better constrain individual SEDs. Ultimately, \textsc{RadioSED} promises to help us better understand the formation, triggering and local environments of radio AGN by working alongside existing IR-Optical-UV SED fitting codes to provide a more holistic, multiwavelength picture of AGN of all ages and evolutionary states.

\section*{Acknowledgements}
The authors wish to thank the anonymous referee for their comments and suggestions which led to many improvements in the final version of this paper.

This research was supported by an Australian Government Research Training Program (RTP) Scholarship. This research was also supported by the Australian Research Council Centre of Excellence for All Sky Astrophysics in 3 Dimensions (ASTRO 3D), through project number CE170100013.

This research made use of the VizieR catalogue access tool, cross-match service XMatch and SIMBAD database provided by CDS, Strasbourg \citep{Ochsenbein2000, Wegner2000}. We also made use of Astroquery \citep{Ginsburg2019Astroquery:Python} and Astropy:\footnote{http://www.astropy.org} a community-developed core Python package and an ecosystem of tools and resources for astronomy \citep{astropy:2013, astropy:2018, astropy:2022}.

This scientific work uses data obtained from Inyarrimanha Ilgari Bundara / the Murchison Radio-astronomy Observatory. We acknowledge the Wajarri Yamaji People as the Traditional Owners and native title holders of the Observatory site. CSIRO’s ASKAP radio telescope is part of the Australia Telescope National Facility\footnote{https://ror.org/05qajvd42}. Operation of ASKAP is funded by the Australian Government with support from the National Collaborative Research Infrastructure Strategy. ASKAP uses the resources of the Pawsey Supercomputing Research Centre. Establishment of ASKAP, Inyarrimanha Ilgari Bundara, the CSIRO Murchison Radio-astronomy Observatory and the Pawsey Supercomputing Research Centre are initiatives of the Australian Government, with support from the Government of Western Australia and the Science and Industry Endowment Fund.

\section*{Data Availability}

All flux density measurements used to construct the SEDs in this work are drawn from publicly available catalogues accessible via the CDS VizieR catalogue service. The code for \textsc{RadioSED}, including crossmatch details for the radio surveys discussed, is accessible on Github and available for use: \url{https://github.com/ekerrison/RadioSED/} (DOI: 10.5281/zenodo.8336846). Data on the individual fits presented in this paper is available upon request to the authors.


\typeout{}
\bibliographystyle{mnras}
\bibliography{references} 




\appendix
\section{SED parameters for a selection of radio sources from the literature}\label{sec:appendix_table}

We present here a table of the SED parameters for a selection of radio sources drawn from the literature, chosen for their representative SED shapes. 

\bgroup
\def\arraystretch{1.5}
\defcitealias{Callingham2017}{C17}
\defcitealias{Stein2021TheFrequencies}{S21}
\defcitealias{Chhetri2013TheCatalogue}{C13}
\defcitealias{Aditya2024}{A24}
\defcitealias{ODea1998}{O98}
\begin{table*}
	\centering
	\caption{SED parameters for sources from the literature fit by each of the four model shapes used in \textsc{RadioSED}. Source names are from the RACS-DR1 catalogue, and literature values for model parameters (peak frequency: $\nu_\text{p, lit}$, and spectral indices: $\alpha_\text{low, lit}$, $\alpha_\text{high, lit}$) are drawn from the citations listed. Note that these typically do not correspond exactly to optically thick and thin spectral indices $\alpha_\text{thick}$, and $\alpha_\text{thin}$ as produced by \textsc{RadioSED}. We have not included a column for peak flux density from the literature since none of the works used as references provide these values. For the exact frequencies at which the literature spectral indices are calculated the reader is directed to the paper in column 12 (Refs), except for sources with parameters derived from the AT20G High Angular Resolution Catalogue \citep{Chhetri2013TheCatalogue}, as that work reports several spectral indices at various intervals. In the case of those sources, we have chosen $\alpha_{low}$ to be the spectral index reported between $1-4.8$\,GHz, and $\alpha_{high}$ to be the spectral index between $19-20$\,GHz. All other model parameters are derived from \textsc{RadioSED}. Citations are not exhaustive for each source, but represent those works in which the source SED shape was first identified, and from which the literature parameters were derived. The citations in Column 11 are as follows: O98: \citet{ODea1998}, C13: \citet{Chhetri2013TheCatalogue}, C17: \citet{Callingham2017}, S21: \citet{Stein2021TheFrequencies}, and A24: \citet{Aditya2024}. Where a model parameter is unknown or not applicable to a particular source, that column is marked with a dash. Sources are ordered by \textsc{RadioSED} peak frequency where available, other\textbf{}wise by $\alpha_\text{thin}$. Those marked with an asterisk are classified are discussed in more detail in-text.}
	\label{tab:obs_sources}
	\begin{tabular}{lccccccccccccccr} 
		\hline
		  Name  & $\nu_\text{p,lit}$ & $\alpha_\text{thick,lit}$ & $\alpha_\text{thin, lit}$ & $\nu_\text{p}$ & $S_\text{p}$ & $\alpha_\text{thick}$ & $\alpha_\text{thin}$ & $\alpha_\text{retrig}$ & Model & Refs.\\
            & (GHz)  & & & (GHz) & (Jy) & & & \\
            \hline
J225350.2-272434 & - & - & -0.60 & - & - & - & $-0.69^{+0.01}_{-0.01}$ & - & \ref{eqn:powlaw} & \citetalias{Stein2021TheFrequencies} \\
J231028.8-105426 & - & - & -0.72 & - & - & - & $-0.70^{+0.01}_{-0.01}$ & - & \ref{eqn:powlaw} & \citetalias{Stein2021TheFrequencies} \\
J224131.5-005144 & - & - & -1.08 & - & - & - & $-0.76^{+0.06}_{-0.05}$ & - & \ref{eqn:powlaw} & \citetalias{Stein2021TheFrequencies} \\
J014808.4-005546 & - & - & -1.11 & - & - & - & $-0.77^{+0.01}_{-0.01}$ & - & \ref{eqn:powlaw} & \citetalias{Stein2021TheFrequencies} \\
J093046.1-220708 & - & - & -1.12 & - & - & - & $-0.83^{+0.01}_{-0.01}$ & - & \ref{eqn:powlaw} & \citetalias{Stein2021TheFrequencies} \\
J231158.6-194733 & - & - & -0.75 & - & - & - & $-0.89^{+0.07}_{-0.07}$ & - & \ref{eqn:powlaw} & \citetalias{Stein2021TheFrequencies} \\
J232850.6-142146 & - & - & -0.89 & - & - & - & $-0.91^{+0.06}_{-0.03}$ & - & \ref{eqn:powlaw} & \citetalias{Stein2021TheFrequencies} \\
J205658.1-123732 & - & - & -0.96 & - & - & - & $-1.07^{+0.06}_{-0.04}$ & - & \ref{eqn:powlaw} & \citetalias{Stein2021TheFrequencies} \\
J040106.2+030057 & 0.325 & $1.71\pm0.31$ & $-0.61\pm0.08$ & $0.33^{+0.02}_{-0.02}$ & $0.34^{+0.01}_{-0.01}$ & $2.37^{+0.28}_{-0.30}$ & $-0.52^{+0.05}_{-0.04}$ & - & \ref{eqn:snellen} & \citetalias{Stein2021TheFrequencies} \\
J203713.4-001057 & 0.325 & $0.73\pm0.32$ & $-0.74\pm0.08$ & $0.41^{+0.03}_{-0.04}$ & $0.86^{+0.03}_{-0.02}$ & $1.01^{+0.12}_{-0.14}$ & $-0.90^{+0.04}_{-0.04}$ & - & \ref{eqn:snellen} & \citetalias{Stein2021TheFrequencies} \\
J231951.4+282829 & 0.325  & $0.4\pm0.18$ & $-1.06\pm 0.07$ & $0.46^{+0.07}_{-0.08}$ & $2.57^{+0.22}_{-0.21}$ & $0.44^{+0.18}_{-0.22}$ & $-1.25^{+0.10}_{-0.10}$ & - & \ref{eqn:snellen} & \citetalias{Stein2021TheFrequencies} \\
J044535.3-001901 & >0.843 & 2.04 & 0.48 & $0.55^{+0.04}_{-0.04}$ & $0.58^{+0.01}_{-0.01}$ & $1.63^{+0.14}_{-0.15}$ & $-0.83^{+0.05}_{-0.04}$ & - & \ref{eqn:snellen} & \citetalias{Callingham2017} \\
J212339.1-011235 & 0.325 & $0.5\pm0.31$ & $-0.61\pm 0.11$ & $0.65^{+0.02}_{-0.03}$ & $1.85^{+0.04}_{-0.04}$ & $0.53^{+0.03}_{-0.03}$ & $-1.10^{+0.02}_{-0.02}$ & - & \ref{eqn:snellen} & \citetalias{Stein2021TheFrequencies} \\
J034039.1+000534 & 0.325 & $0.41\pm0.32$ & $-0.53\pm 0.11$ & $0.71^{+0.15}_{-0.18}$ & $0.22^{+0.03}_{-0.03}$ & $0.28^{+0.11}_{-0.14}$ & $-1.01^{+0.12}_{-0.12}$ & - & \ref{eqn:snellen} & \citetalias{Stein2021TheFrequencies} \\
J002225.4+001456 & 1.4 & $0.31\pm0.08$ & $-0.82\pm 0.10$ & $0.81^{+0.02}_{-0.02}$ & $4.22^{+0.04}_{-0.04}$ & $1.06^{+0.03}_{-0.03}$ & $-1.15^{+0.01}_{-0.01}$ & - & \ref{eqn:snellen} & \citetalias{Stein2021TheFrequencies} \\
J161511.5-243059 & 1.4 & $1.29\pm0.12$ & $-0.80\pm 0.23$ & $1.07^{+0.03}_{-0.03}$ & $0.94^{+0.01}_{-0.02}$ & $1.82^{+0.06}_{-0.06}$ & $-1.11^{+0.02}_{-0.02}$ & - & \ref{eqn:snellen} & \citetalias{Stein2021TheFrequencies} \\
J002604.7-475617 & 0.213 & -0.99 & 1.20 & $0.24^{+0.01}_{-0.01}$ & $0.99^{+0.01}_{-0.01}$ & $1.45^{+0.06}_{-0.06}$ & $-1.62^{+0.03}_{-0.03}$ & - & \ref{eqn:orienti} & \citetalias{Callingham2017} \\
*J230659.4-652132 & - & 0.01 & -1.11 & $1.55^{+0.04}_{-0.04}$ & $0.43^{+0.01}_{-0.01}$ & $1.23^{+0.04}_{-0.04}$ & $-1.90^{+0.04}_{-0.03}$ & - & \ref{eqn:orienti} & \citetalias{Chhetri2013TheCatalogue} \\
J234708.6-185618 & - & -0.13 & -0.99 & $1.82^{+0.07}_{-0.06}$ & $0.60^{+0.01}_{-0.01}$ & $1.25^{+0.07}_{-0.07}$ & $-1.76^{+0.07}_{-0.07}$ & - & \ref{eqn:orienti} & \citetalias{Chhetri2013TheCatalogue} \\
J002442.8-420202 & - & 0.06 & -1.48 & $2.03^{+0.01}_{-0.01}$ & $2.62^{+0.01}_{-0.01}$ & $2.55^{+0.04}_{-0.04}$ & $-3.43^{+0.05}_{-0.05}$ & - & \ref{eqn:orienti} & \citetalias{Chhetri2013TheCatalogue} \\
*J233159.4-381146 & - & 0.07 & -0.43 & $2.85^{+0.15}_{-0.15}$ & $0.50^{+0.01}_{-0.01}$ & $0.72^{+0.04}_{-0.04}$ & $-0.92^{+0.04}_{-0.04}$ & - & \ref{eqn:orienti} & \citetalias{Chhetri2013TheCatalogue} \\
*J231359.7-370445 & - & 0.30 & -0.51 & $3.69^{+0.13}_{-0.13}$ & $0.29^{+0.01}_{-0.01}$ & $0.81^{+0.03}_{-0.03}$ & $-0.75^{+0.04}_{-0.03}$ & - & \ref{eqn:orienti} & \citetalias{Chhetri2013TheCatalogue} \\
*J230738.6-224753 & - & 0.60 & -0.61 & $5.80^{+0.12}_{-0.11}$ & $0.71^{+0.02}_{-0.02}$ & $1.21^{+0.03}_{-0.03}$ & $-1.82^{+0.06}_{-0.06}$ & - & \ref{eqn:orienti} & \citetalias{Chhetri2013TheCatalogue} \\
*J012528.8-000555 & >0.843 & 0.46 & 0.27 & $12.69^{+0.81}_{-0.84}$ & $1.59^{+0.02}_{-0.03}$ & $0.26^{+0.01}_{-0.01}$ & $-0.77^{+0.02}_{-0.02}$ & - & \ref{eqn:orienti} &  \citetalias{Callingham2017} \\
J053354.5-234429 & >0.843 & 0.71 & 0.29 & $0.63^{+0.01}_{-0.01}$ & $1.82^{+0.03}_{-0.03}$ & $0.84^{+0.03}_{-0.02}$ & $-0.77^{+0.02}_{-0.02}$ & $-0.37^{+0.05}_{-0.04}$ & \ref{eqn:retrig} & \citetalias{Callingham2017} \\
J010551.0+155342 & 1.4 & $0.64\pm0.1$ & $-0.75\pm0.14$ & $0.70^{+0.02}_{-0.02}$ & $1.40^{+0.02}_{-0.02}$ & $1.38^{+0.03}_{-0.03}$ & $-0.52^{+0.09}_{-0.08}$ & $-0.66^{+0.47}_{-0.30}$ & \ref{eqn:retrig} & \citetalias{Stein2021TheFrequencies} \\
*J015310.1-331025 & - & -0.43 & 0.38 & $0.95^{+0.05}_{-0.03}$ & $1.20^{+0.01}_{-0.01}$ & $0.99^{+0.10}_{-0.06}$ & $-0.32^{+0.01}_{-0.01}$ & $-0.50^{+0.09}_{-0.06}$ & \ref{eqn:retrig} & \citetalias{Callingham2017} \\
J044133.7-334004 & - & -0.48 & 0.75 & $1.49^{+0.02}_{-0.02}$ & $1.14^{+0.01}_{-0.01}$ & $0.71^{+0.04}_{-0.03}$ & $-0.86^{+0.01}_{-0.01}$ & $-2.29^{+0.18}_{-0.28}$ & \ref{eqn:retrig} & \citetalias{Callingham2017} \\
*J135706.0-174401 & - & -0.27 & 0.25 & $1.93^{+0.06}_{-0.06}$ & $1.50^{+0.03}_{-0.03}$ & $0.28^{+0.02}_{-0.02}$ & $-0.87^{+0.02}_{-0.02}$ & $-1.15^{+0.30}_{-0.18}$ & \ref{eqn:retrig} & \citetalias{Callingham2017} \\
*J212912.1-153840 & - & -0.51 & 0.34 & $6.93^{+0.22}_{-0.23}$ & $1.52^{+0.06}_{-0.07}$ & $0.67^{+0.01}_{-0.01}$ & $-0.77^{+0.11}_{-0.10}$ & $-0.70^{+0.08}_{-0.08}$ & \ref{eqn:retrig} & \citetalias{Callingham2017} \\
*J213638.5+004154 & 4.3 & - & - & $7.73^{+0.33}_{-0.26}$ & $9.76^{+0.28}_{-0.31}$ & $0.75^{+0.01}_{-0.01}$ & $-0.85^{+0.02}_{-0.02}$ & $-1.36^{+0.08}_{-0.08}$ & \ref{eqn:retrig} & \citetalias{ODea1998} \\
*J225805.9-275821 & - & - & - & $61.82^{+0.01}_{-0.01}$ & $5.26^{+0.03}_{-0.03}$ & $0.39^{+0.01}_{-0.01}$ & $-0.57^{+0.01}_{-0.01}$ & $-0.94^{+0.02}_{-0.02}$ & \ref{eqn:retrig} & \citetalias{Aditya2024} \\
		\hline
	\end{tabular}
\end{table*}
\egroup


\bsp	
\label{lastpage}
\end{document}